# The capacity region of a product of two unmatched Gaussian broadcast channels with three particular messages and a common message

Ramy H. Gohary and Timothy N. Davidson

**Abstract**

This paper considers a Gaussian broadcast channel with two unmatched degraded components, three particular messages, and a common message that is intended for all three receivers. It is shown that for this channel superposition coding and Gaussian signalling is sufficient to achieve every point in the capacity region.

## I. Introduction

In a broadcast (BC) channel a single transmitter sends messages to multiple receivers [1]. These messages may be common to all receivers, or particular to an individual receiver or a subset of receivers. The vector containing the rates of these messages is said to be achievable if each receiver is able to reliably decode its intended messages. The closure of all such vectors is usually referred to as the capacity region [2].

The first author is with the Communications Research Centre, Ottawa, Ontario, Canada, ramy.gohary@crc.gc.ca. The second author is with The Department of Electrical and Computer Engineering, McMaster University, Hamilton, Ontario, Canada, davidson@mcmaster.ca.

A special class of BC channels is the one in which the received signals form a Markov chain. In this case, the received signals are said to be degraded versions of each other, and the degradation level of each signal is given by its order in the Markov chain. For the class of degraded channels, superposition coding [3] is known to attain every point on the boundary of the capacity region in the general unrestricted case [4], and in the case of Gaussian channels with a power constraint [5].

If the received signals do not form a Markov chain, the BC channel is said to be non-degraded, and the coding scheme developed in [3] does not apply directly [6]. Although degraded channels are useful in modelling single-input single-output BC systems, many practical systems give rise to non-degraded channels, including those that employ multicarrier transmission [7], and the class of multiple-input multiple-output (MIMO) systems [6], [8].

Most of the studies on non-degraded BC channels have focused on the case in which only independent particular messages are sent to the receivers, e.g., [6], [8]–[16]. For example, the sum capacity for the case in which particular messages are broadcast over Gaussian MIMO channels was studied in [15] and [16] and was shown in [8], [12], [13] to be achievable by dirty paper coding (DPC) [17] with Gaussian signalling. Later, it was shown in [6] that DPC with Gaussian signalling is sufficient to attain every point in the achievable rate region. That is, DPC with Gaussian signalling is sufficient for achieving every point in the capacity region of the Gaussian MIMO BC channel with particular messages.

In contrast to the case of particular messages only, there has been less progress in characterizing the capacity region of general non-degraded BC channels when common or partially common messages are to be transmitted along with particular messages. However, some partial results are available. For instance, for the case in which common messages may be transmitted over general non-degraded BC channels, characterizations of achievable inner bounds were obtained in [18], [19] and [20], and characterizations of outer bounds were obtained in [21]. In addition, characterizations of the capacity region of a BC channel with two receivers, two unmatched parallel degraded components, a common message intended for both receivers and a particular message intended for each receiver were provided in [22]. For a BC channel with three receivers, a common message and one particular message, a single-letter characterization of the capacity region was provided in [23] and this region was shown to be strictly larger than the one conjectured in [24]. For general BC channels in which common, partially common and particular messages are intended for the receivers, fundamental constraints on the geometry of the capacity region were provided in [25].

In this paper we consider a different class of BC channels with three receivers. In contrast to [23], in which there is only one particular message, in the class considered herein a particular message is sent to each of the three receivers, in addition to the common message. The channel is assumed to be Gaussian and memoryless with two unmatched degraded components. It will be shown that for the degradation orders considered in this paper, superposition coding and Gaussian signalling are sufficient to attain any point on the boundary of the capacity region.

Our methodology for obtaining this result involves three stages. First, we provide an ostensibly relaxed characterization of the rate region that can be attained by superposition coding and Gaussian signalling. Using the Karush-Kuhn-Tucker (KKT) optimality conditions, this relaxation is shown to be tight. Second, we use information-theoretic analysis to obtain bounds on any achievable rate vector. Finally, by combining the tight relaxation and the information-theoretic bounds, we establish the desired converse, i.e., that every achievable rate vector can be attained by superposition coding and Gaussian signalling.

The paper is organized as follows. In Section II we describe the system model considered in this paper along with necessary definitions and notations. In Section III we provide a characterization of the rate region that can be achieved by superposition coding and Gaussian signalling, and in Section IV we express the boundary of this region as an optimization problem. In Section V we consider a relaxation of the optimization problem in Section IV. In Section VI we provide common solutions for the KKT conditions that correspond to both the original and the relaxed optimization problems. Using these solutions, in Section VII we establish the tightness of the relaxation in Section V. In Section VIII, we obtain information theoretic bounds on the achievable rates and in Section IX, we employ the entropy power inequality to the bounds of Section VIII. The resulting inequalities are then identified with the relaxed characterization of the superposition coding and Gaussian signalling rate region developed in Section V. This identification is then used in Section IX to establish the main result of the paper. That is, the optimality of superposition coding and Gaussian signalling. Section X concludes the paper. For clarity of exposition, most of the proofs are relegated to the appendices.

*Notation:* The paper uses conventional notation throughout. Vectors are denoted by regular weight symbols, and subscripts and superscripts are used to refer to particular entries of these vectors.

## II. System Model

We consider the discrete-time BC channel depicted in Figure 1. For this channel the transmitter sends messages to three receivers over two parallel unmatched Gaussian memoryless degraded subchannels. The transmitted signal on subchannel $i$ is denoted by $X_i$ and its power is denoted by $P_i$. The signal observed by receivers $Y$, $Z$ and $W$ on the $i$-th subchannel is denoted by $Y_i$, $Z_i$ and $W_i$, $i = 1, 2$, respectively.

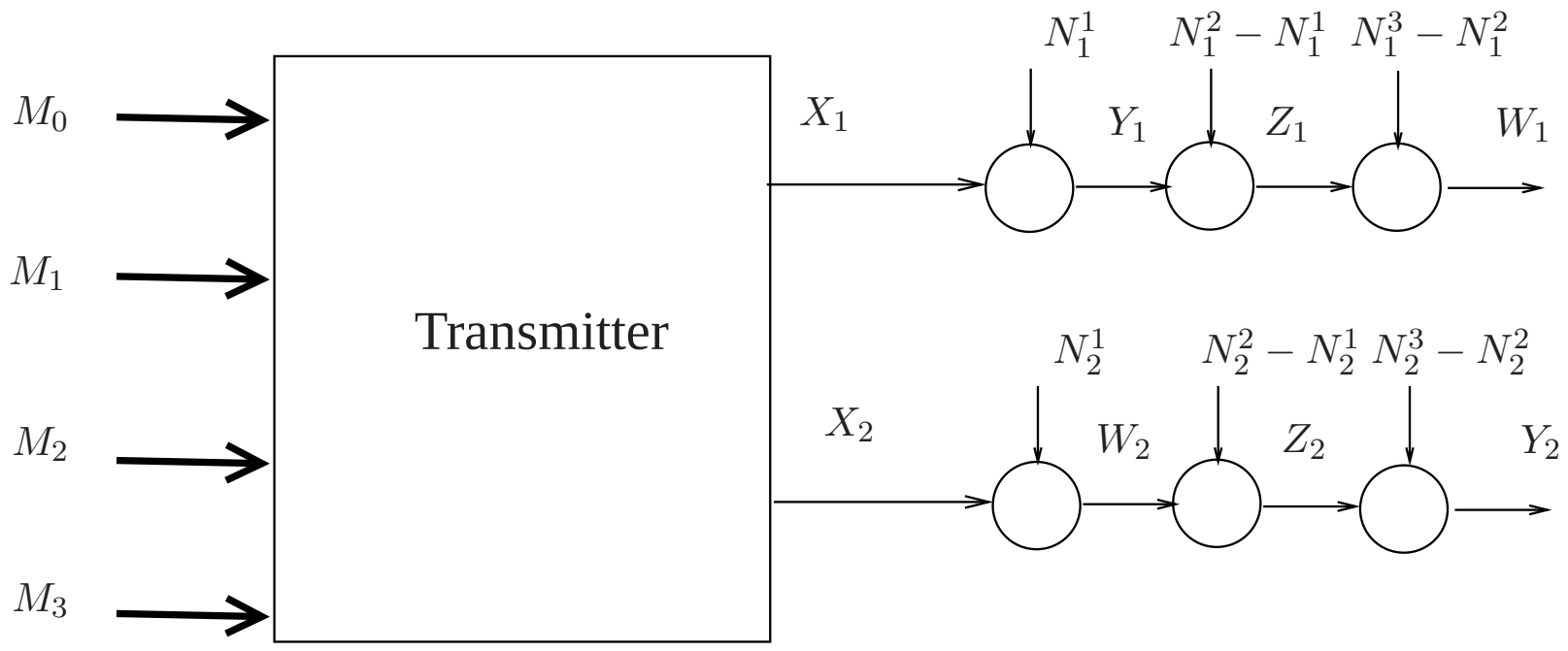

Fig. 1. A BC channel with 2 unmatched degraded components and 3 receivers.

The variance of the Gaussian noise at degradation level $j$ on subchannel $i$ is denoted by $N_i^j$, where $N_i^j < N_i^{j+1}$, $i, j = 1, 2$. The transmitter wishes to send particular messages of rates $R_1$, $R_2$ and $R_3$ to receivers $Y$, $Z$, and $W$, respectively, and also wishes to send a common message of rate $R_0$ to all three receivers. We will show that, for the scenario in Figure 1, the region of rate vectors $(R_0, R_1, R_2, R_3)$ that are achievable using superposition coding and Gaussian signalling is the region of all achievable rates, i.e., the capacity region.

The description of the set of rates that are achievable by superposition coding (SPC) is parameterized by a set of power partitions, cf. [22]. Each partition determines the fraction of the transmission power that is used to transmit the incremental message to the next less degraded receiver on each sub-channel. Since power partitions on each subchannel lie in the unit simplex in $\mathbb{R}^3$ we need only specify two of them on each subchannel. Given this vector of partitions, $\alpha = [\alpha_1^1, \alpha_1^2, \alpha_2^1, \alpha_2^2]$, where $\alpha_i^j \geq 0$ and $\alpha_i^1 + \alpha_i^2 \leq 1$, the following definitions will simplify the description of the SPC achievable rate region

$$f_0(\alpha) \triangleq \frac{1}{2}\log\Big(\frac{N_1^1 + P_1}{N_1^1 + (\alpha_1^1 + \alpha_1^2)P_1}\Big) + \frac{1}{2}\log\Big(\frac{N_2^3 + P_2}{N_2^3 + (\alpha_2^1 + \alpha_2^2)P_2}\Big), \tag{1a}$$

$$f_{01}(\alpha) \triangleq \frac{1}{2}\log\Big(\frac{N_1^1 + P_1}{N_1^1}\Big) + \frac{1}{2}\log\Big(\frac{N_2^3 + P_2}{N_2^3 + (\alpha_2^1 + \alpha_2^2)P_2}\Big), \tag{1b}$$

$$f_{012}(\alpha) \triangleq f_{01}(\alpha) + \frac{1}{2}\log\Big(\frac{N_2^2 + (\alpha_2^1 + \alpha_2^2)P_2}{N_2^2 + \alpha_2^1 P_2}\Big), \tag{1c}$$

$$f_{0123}(\alpha) \triangleq f_{012}(\alpha) + \frac{1}{2}\log\Big(\frac{N_2^1 + \alpha_2^1 P_2}{N_2^1}\Big), \tag{1d}$$

$$g_0(\alpha) \triangleq \frac{1}{2}\log\Big(\frac{N_1^2 + P_1}{N_1^2 + (\alpha_1^1 + \alpha_1^2)P_1}\Big) + \frac{1}{2}\log\Big(\frac{N_2^2 + P_2}{N_2^2 + (\alpha_2^1 + \alpha_2^2)P_2}\Big), \tag{1e}$$

$$g_{02}(\alpha) \triangleq \frac{1}{2}\log\Big(\frac{N_1^2 + P_1}{N_1^2 + \alpha_1^1 P_1}\Big) + \frac{1}{2}\log\Big(\frac{N_2^2 + P_2}{N_2^2 + \alpha_2^1 P_2}\Big), \tag{1f}$$

$$g_{012}(\alpha) \triangleq g_{02}(\alpha) + \frac{1}{2}\log\Big(\frac{N_1^1 + \alpha_1^1 P_1}{N_1^1}\Big), \tag{1g}$$

$$g_{023}(\alpha) \triangleq g_{02}(\alpha) + \frac{1}{2}\log\Big(\frac{N_2^1 + \alpha_2^1 P_2}{N_2^1}\Big), \tag{1h}$$

$$g_{0123}(\alpha) \triangleq g_{012}(\alpha) + \frac{1}{2}\log\Big(\frac{N_2^1 + \alpha_2^1 P_2}{N_2^1}\Big) = g_{023}(\alpha) + \frac{1}{2}\log\Big(\frac{N_1^1 + \alpha_1^1 P_1}{N_1^1}\Big) \tag{1i}$$

$$h_0(\alpha) \triangleq \frac{1}{2}\log\Big(\frac{N_1^3 + P_1}{N_1^3 + (\alpha_1^1 + \alpha_1^2)P_1}\Big) + \frac{1}{2}\log\Big(\frac{N_2^1 + P_2}{N_2^1 + (\alpha_2^1 + \alpha_2^2)P_2}\Big), \tag{1j}$$

$$h_{03}(\alpha) \triangleq \frac{1}{2}\log\Big(\frac{N_1^3 + P_1}{N_1^3 + (\alpha_1^1 + \alpha_1^2)P_1}\Big) + \frac{1}{2}\log\Big(\frac{N_2^1 + P_2}{N_2^1}\Big), \tag{1k}$$

$$h_{023}(\alpha) \triangleq h_{03}(\alpha) + \frac{1}{2}\log\Big(\frac{N_1^2 + (\alpha_1^1 + \alpha_1^2)P_1}{N_1^2 + \alpha_1^1 P_1}\Big), \tag{1l}$$

$$h_{0123}(\alpha) \triangleq h_{023}(\alpha) + \frac{1}{2}\log\Big(\frac{N_1^1 + \alpha_1^1 P_1}{N_1^1}\Big). \tag{1m}$$

## III. An achievable Rate Region

In this section we will characterize the set of rates that are achievable by superposition coding (SPC) for the BC channel depicted in Figure 1.

First, let us consider the case in which the transmit powers $P_1$ and $P_2$ are given. By applying the fundamental principles of superposition coding [1], [3], and by coding the common message jointly over the sub-channels [22] and coding the components of the particular messages that are transmitted on each sub-channel separately [22], it can be shown that a rate vector $R$ can be achieved using superposition coding (and Gaussian signalling) with the power partitions $\theta$ if the following inequalities are satisfied:

$$R_0 \leq f_0(\theta), \quad R_0 + R_1 \leq f_{01}(\theta), \quad R_0 + R_1 + R_2 \leq f_{012}(\theta), \tag{2a}$$

$$R_0 + R_1 + R_2 + R_3 \leq f_{0123}(\theta), \tag{2b}$$

$$R_0 \leq g_0(\theta), \quad R_0 + R_2 \leq g_{02}(\theta), \tag{2c}$$

$$R_0 + R_1 + R_2 \leq g_{012}(\theta), \tag{2d}$$

$$R_0 + R_2 + R_3 \leq g_{023}(\theta), \tag{2e}$$

$$R_0 + R_1 + R_2 + R_3 \leq g_{0123}(\theta), \tag{2f}$$

$$R_0 \leq h_0(\theta), \quad R_0 + R_3 \leq h_{03}(\theta), \quad R_0 + R_2 + R_3 \leq h_{023}(\theta), \tag{2g}$$

$$R_0 + R_1 + R_2 + R_3 \leq h_{0123}(\theta), \tag{2h}$$

$$R_k \geq 0, \quad k = 0, 1, 2, 3, \tag{2i}$$

$$\theta \in \bar{\mathcal{S}}, \tag{2j}$$

where

$$\bar{\mathcal{S}} \triangleq \{\theta | \theta_i^\ell \geq 0,\ \theta_i^1 + \theta_i^2 \leq 1,\ i = 1, 2\}. \tag{3}$$

In the derivation of the constraints in (2) we have used the fact that for the BC channel shown in Figure 1, the constraints

$$R_0 + R_1 + R_3 \leq f_{01}(\theta) + \frac{1}{2}\log\Big(\frac{N_2^1 + (\theta_2^2 + \theta_2^1)P_2}{N_2^1}\Big), \tag{4a}$$

$$R_0 + R_1 + R_3 \leq h_{03}(\theta) + \frac{1}{2}\log\Big(\frac{N_1^1 + (\theta_1^2 + \theta_1^1)P_1}{N_1^1}\Big) \tag{4b}$$

are redundant. To show that (4a) is redundant, we observe that

$$\begin{aligned} f_{01}(\theta) + \frac{1}{2}\log\Big(\frac{N_2^1 + (\theta_2^2 + \theta_2^1)P_2}{N_2^1}\Big) &= f_{01}(\theta) + \frac{1}{2}\log\Big(\frac{N_2^1 + (\theta_2^1 + \theta_2^2)P_2}{N_2^1 + \theta_2^1 P_2}\Big) + \frac{1}{2}\log\Big(\frac{N_2^1 + \theta_2^1 P_2}{N_2^1}\Big) \\ &> f_{01}(\theta) + \frac{1}{2}\log\Big(\frac{N_2^2 + (\theta_2^1 + \theta_2^2)P_2}{N_2^2 + \theta_2^1 P_2}\Big) + \frac{1}{2}\log\Big(\frac{N_2^1 + \theta_2^1 P_2}{N_2^1}\Big) \quad (5) \\ &= f_{0123}(\theta), \quad (6) \end{aligned}$$

where in (5) we have used the fact that $\log\Big(\frac{N_2^1+(\theta_2^1+\theta_2^2)P_2}{N_2^1+\theta_2^1 P_2}\Big)$ is monotonically decreasing in $N_2^1$, and that $N_2^1 < N_2^2$. Now, for any $R_2 > 0$, the left hand side of (4a) is strictly less than the left hand side of the inequality in (2b) and the right hand side of (4a) is strictly greater than the right hand side of the inequality in (2b). Hence, we conclude that the last inequality in (2b) is strictly tighter than the inequality in (4a), and hence the redundancy of (4a). A similar argument involving (2h) can be used to show that (4b) is also redundant.

To facilitate the analysis of the rates and partitions that satisfy the inequalities in (2), we will expurgate a set of partitions, $\mathcal{S}_N$, that do not yield rates on the boundary of the region described by (2). The set of rates on the boundary of the region characterized by (2) can be expressed as the set of rates generated by the optimization problem

$$\max_{R,\theta} \sum_{k=0}^{3} v_k R_k, \quad \text{subject to the constraints in (2)}, \tag{7}$$

for all non-negative weight vectors $v \in \mathbb{R}^4$. We will refer to the rate vector generated by (7) for a certain power partition $\theta$ and a weight vector $v$, as $R^*(\theta, v)$. The set $\mathcal{S}_N$ of partitions that do not solve (7) for any weight vector $v$ can be written as

$$\mathcal{S}_N \triangleq \Big\{\theta | \forall v \in \mathbb{R}^4, \exists \vartheta \in \bar{\mathcal{S}} \text{ such that } \sum_{k=0}^{3} v_k R_k^*(\theta, v) < \sum_{k=0}^{3} v_k R_k^*(\vartheta, v)\Big\}. \tag{8}$$

By definition, excluding $\mathcal{S}_N$ from the set of feasible power partitions in (3) does not affect the set of rates that are achievable by superposition coding and Gaussian signalling.

We now define equivalence classes of power partitions. In particular, the partition $\theta$ is said to be equivalent to the partition $\vartheta$ with respect to $v$ if and only if $R^*(\theta, v) = R^*(\vartheta, v)$. The set of partitions that are equivalent to $\theta$ for a given weight vector, $v$, can be denoted by $[\theta]_v$ and can be represented by $\theta_v$. If we restrict the set of feasible partitions to the quotient space $(\bar{\mathcal{S}} \setminus \mathcal{S}_N)/\!\sim$, the set of rates that are achievable by superposition coding and Gaussian signalling remains the same as that generated by the constraints in (2) for all $\theta \in \bar{\mathcal{S}}$. Furthermore, restricting our attention to partitions in

$$\mathcal{S} \triangleq \Big\{\theta_v \Big| v \in \mathbb{R}^4\Big\} \tag{9}$$

yields the following result, which we will use in Section IV.

*Lemma 1:* The restriction of the power partitions to $\mathcal{S}$ in (9) rather than $\bar{\mathcal{S}}$ in (3) establishes a one-to-one correspondence between points on the boundary of the rate region that can be achieved by superposition coding and Gaussian signalling and the power partitions that enable these rates to be achieved. □

Now, let us define the SPC-Region for a given power allocation $(P_1, P_2)$ to be the region containing partition-rate vectors $(\theta, R)$ such that the rate vector $R$ is achievable using superposition coding with the power partitions specified by $\theta$. More precisely, let[1]

$$\text{SPC-Region}(P_1, P_2) = \Big\{(\theta, R) \,|\, \text{constraints (2a)–(2i) satisfied}, \theta \in \mathcal{S}\Big\}. \tag{10}$$

The association of partitions with rates is introduced to enable us to express the SPC-Region as the intersection of two regions, namely, Region$_1$ and Region$_2$, below. For a system in which the transmission powers $P_1$ and $P_2$ may be allocated arbitrarily, subject to a total power constraint of the form $P_1 + P_2 \leq \bar{P}$, the complete SPC-Region is

$$\bigcup_{\substack{P_1, P_2 \geq 0,\\ P_1 + P_2 \leq \bar{P}}} \text{SPC-Region}(P_1, P_2). \tag{11}$$

Now that we have a characterization of the SPC-region, the rest of the paper is focused on establishing the converse. Given the structure of the expression in (11), to establish the converse it is sufficient to consider each power allocation $(P_1, P_2)$ separately, and to show that for every rate vector $R$ that is achievable with a given power allocation there exists a vector of power partitions, $\theta$, such that the inequalities in (2) are satisfied, i.e., such that $(\theta, R) \in \text{SPC-Region}(P_1, P_2)$. Since we will deal with each power allocation separately, we will simplify our notation and drop the explicit dependence of the SPC-Region on $P_1$ and $P_2$.

[1]Note that the case in which either $P_1$ or $P_2$ is zero corresponds to a degraded channel case for which the entire rate region has been fully characterized [2]. Therefore, henceforth we will assume that both $P_1$ and $P_2$ are strictly greater than zero.

In our development of the converse, we will show that for any rate vector $R$ that is achievable with transmission powers $P_1$ and $P_2$ there exist power partitions $\theta'$ and $\theta''$ such that

$$(\theta', R) \in \text{Region}_1 \triangleq \big\{(\theta, R) \,|\, \text{constraints (2a)–(2i)} \setminus \text{(2e) satisfied}, \theta \in \mathcal{S}\big\}, \tag{12a}$$

$$(\theta'', R) \in \text{Region}_2 \triangleq \big\{(\theta, R) \,|\, \text{constraints (2a)–(2i)} \setminus \text{(2d) satisfied}, \theta \in \mathcal{S}\big\}, \tag{12b}$$

where $\setminus$ denotes the removal of a constraint. Hence, $\text{Region}_1$ contains all the achievable rates, as does $\text{Region}_2$, and thus so does their intersection. This intersection can be written as

$$\text{Region}_1 \bigcap \text{Region}_2 = \left\{(\theta, R) \,\middle|\, \text{constraints (2a)–(2i)} \setminus \text{(2e)} \bigcup \text{ (2a)–(2i)} \setminus \text{(2d) satisfied }, \theta \in \mathcal{S}\right\} \tag{13a}$$

$$= \big\{(\theta, R) \,|\, \text{constraints (2a)–(2i) satisfied}, \theta \in \mathcal{S}\big\}, \tag{13b}$$

which is the SPC-region in (10). In the remainder of the paper we will focus on showing that every achievable rate vector is contained in $\text{Region}_1$. By exploiting symmetry, a similar argument can be used to show that $\text{Region}_2$ also contains every achievable rate vector, and we will provide some of the details of that proof as they arise.

## IV. The boundary of the SPC-Region as an optimization problem

Lemma 1 implies that points on the boundary of the (partition-rate) SPC-Region can be found by maximizing weighted-sum-rates and associating with each rate vector a partition that achieves it. Using (13), we will find the boundary of the SPC-Region by considering the boundaries of $\text{Region}_1$ and $\text{Region}_2$ separately. First, we observe [22], [26] that any achievable common information rate, $R_0$, must satisfy

$$R_0 \leq R_{0,\max} \triangleq \min\left\{C\Big(\frac{P_1}{N_1^1}\Big) + C\Big(\frac{P_2}{N_2^3}\Big), C\Big(\frac{P_1}{N_1^2}\Big) + C\Big(\frac{P_2}{N_2^2}\Big), C\Big(\frac{P_1}{N_1^3}\Big) + C\Big(\frac{P_2}{N_2^1}\Big)\right\}, \tag{14}$$

where $C(x) \triangleq \frac{1}{2}\log(1+x)$. The arguments of the minimization in (14) are the maximum rates that can be communicated to receivers $Y$, $Z$ and $W$, respectively.

### A. Characterizations of $\text{Region}_1$ and $\text{Region}_2$

Let $R_0$ be a common information rate that satisfies (14) and let $\theta'$ and $\theta''$ be the power partitions that are used to characterize $\text{Region}_1$ and $\text{Region}_2$ in (12a) and (12b), respectively. Consider the partition-rate regions with the following boundaries. In particular, for $\text{Region}_1$ we consider

$$\bigcup_{R_0\in[0,R_{0,\max}]}\bigcup_{w_i\geq 0}\Bigg\{(\theta',R_0,R_1,R_2,R_3)\Bigg|(\theta',R_1,R_2,R_3)=\arg\max_{\{R_k\}_{k=1}^3,\theta'}\sum_{k=1}^{3}w_kR_k,\quad \text{subject to}\quad \text{(2a)–(2i)}\setminus\text{(2e)},\ \theta'\in\bar{\mathcal{S}}\Bigg\},\tag{15}$$

and for $\text{Region}_2$ we consider

$$\bigcup_{R_0\in[0,R_{0,\max}]}\bigcup_{w_i\geq 0}\Bigg\{(\theta'',R_0,R_1,R_2,R_3)\Bigg|(\theta'',R_1,R_2,R_3)=\arg\max_{\{R_k\}_{k=1}^3,\theta''}\sum_{k=1}^{3}w_kR_k,\quad \text{subject to}\quad \text{(2a)–(2i)}\setminus\text{(2d)},\ \theta''\in\bar{\mathcal{S}}\Bigg\}.\tag{16}$$

*Remark 1:* The (partition-rate) SPC-Region can be constructed from the intersection of the region bounded by (15) and that bounded by (16). In particular, the boundary of this region can be constructed in the following way. Suppose that for a given $R_0$ and given $\{w_k\}_{k=1}^3$ the partition-rate vector generated by the problem in (15) is feasible for the problem in (16). Then the partition $\theta'$ is a representative of an equivalence class of partitions that maximize the weighted sum-rate, and we can choose this partition to be the representative in $\mathcal{S}$. If, however, the partition-rate vector generated by (15) does not belong to the feasible set of (16), then taking the intersection of the region bounded by (15) and the region bounded by (16), analogous to (13), eliminates such a vector. A particular instance in which we implement this intersection will be considered in Remark 2 below. □

Given the observation in the above remark, we will henceforth refer to the region bounded by (15) as $\text{Region}_1$ and that bounded by (16) as $\text{Region}_2$.

### *B. KKT necessary optimality conditions for* (15)

In this section we will analyze the characterization of $\text{Region}_1$ in (15). The corresponding analysis for $\text{Region}_2$ follows an analogous path that exploits the symmetry between receivers $Y$ and $W$ in Figure 1.

In order to expose the structure of $\text{Region}_1$ we consider the KKT conditions (cf. [27]) that correspond to the optimization problem in (15). In doing so, we will seek solutions for which $R > 0$, $\theta > 0$ and $\theta_i^1 + \theta_i^2 < 1$, $i = 1, 2$. Due to the continuity of the rate functions, these assumptions are not restrictive because $R$ and $\theta$ can take on arbitrarily small values. Observe that it is not known whether the optimization problem in (15) is convex. Hence, for any regular point of the feasible set of (15), the KKT conditions are only necessary for optimality [27]. Using the definitions in (1), and using $L_0$ to denote the Lagrangian corresponding to (15), we have

$$\frac{\partial L_0}{\partial R_1} = w_1 - (\beta_1 + \beta_2 + \beta_3) - (\beta_6 + \beta_7) - \beta_{11} = 0, \tag{17a}$$

$$\frac{\partial L_0}{\partial R_2} = w_2 - (\beta_2 + \beta_3) - (\beta_5 + \beta_6 + \beta_7) - (\beta_{10} + \beta_{11}) = 0, \tag{17b}$$

$$\frac{\partial L_0}{\partial R_3} = w_3 - \beta_3 - \beta_7 - (\beta_9 + \beta_{10} + \beta_{11}) = 0, \tag{17c}$$

$$\frac{\partial L_0}{\partial \theta_2^1} = -\frac{(\beta_0 + \beta_1 + \beta_2 + \beta_3)P_2}{N_2^3 + (\theta_2^1 + \theta_2^2)P_2} + \frac{(\beta_2 + \beta_3 - \beta_4)P_2}{N_2^2 + (\theta_2^1 + \theta_2^2)P_2} - \frac{\beta_8 P_2}{N_2^1 + (\theta_2^1 + \theta_2^2)P_2} - \frac{(\beta_2 + \beta_3 + \beta_5 + \beta_6 + \beta_7)P_2}{N_2^2 + \theta_2^1 P_2} + \frac{(\beta_3 + \beta_7)P_2}{N_2^1 + \theta_2^1 P_2} = 0, \tag{17d}$$

$$\frac{\partial L_0}{\partial \theta_2^2} = -\frac{(\beta_0 + \beta_1 + \beta_2 + \beta_3)P_2}{N_2^3 + (\theta_2^1 + \theta_2^2)P_2} + \frac{(\beta_2 + \beta_3 - \beta_4)P_2}{N_2^2 + (\theta_2^1 + \theta_2^2)P_2} - \frac{\beta_8 P_2}{N_2^1 + (\theta_2^1 + \theta_2^2)P_2} = 0, \tag{17e}$$

$$\frac{\partial L_0}{\partial \theta_1^2} = \frac{-\beta_0 P_1}{N_1^1 + (\theta_1^1 + \theta_1^2)P_1} + \frac{(\beta_{10} + \beta_{11} - \beta_4)P_1}{N_1^2 + (\theta_1^1 + \theta_1^2)P_1} - \frac{(\beta_8 + \beta_9 + \beta_{10} + \beta_{11})P_1}{N_1^3 + (\theta_1^1 + \theta_1^2)P_1} = 0, \tag{17f}$$

$$\frac{\partial L_0}{\partial \theta_1^1} = \frac{-\beta_0 P_1}{N_1^1 + (\theta_1^1 + \theta_1^2)P_1} + \frac{(\beta_{10} + \beta_{11} - \beta_4)P_1}{N_1^2 + (\theta_1^1 + \theta_1^2)P_1} - \frac{(\beta_8 + \beta_9 + \beta_{10} + \beta_{11})P_1}{N_1^3 + (\theta_1^1 + \theta_1^2)P_1} - \frac{(\beta_5 + \beta_6 + \beta_7 + \beta_{10} + \beta_{11})P_1}{N_1^2 + \theta_1^1 P_1} + \frac{(\beta_6 + \beta_7 + \beta_{11})P_1}{N_1^1 + \theta_1^1 P_1} = 0, \tag{17g}$$

$$\beta_0 \geq 0, \quad f_0(\theta) \geq R_0, \quad \beta_0(f_0(\theta) - R_0) = 0, \tag{17h}$$

$$\beta_1 \geq 0, \quad f_{01}(\theta) \geq R_0 + R_1, \quad \beta_1(f_{01}(\theta) - R_0 - R_1) = 0, \tag{17i}$$

$$\beta_2 \geq 0, \quad f_{012}(\theta) \geq R_0 + \sum_{k=1}^{2} R_k, \quad \beta_2\big(f_{012}(\theta) - R_0 - \sum_{k=1}^{2} R_k\big) = 0, \tag{17j}$$

$$\beta_3 \geq 0, \quad f_{0123}(\theta) \geq R_0 + \sum_{k=1}^{3} R_k, \quad \beta_3\big(f_{0123}(\theta) - R_0 - \sum_{k=1}^{3} R_k\big) = 0, \tag{17k}$$

$$\beta_4 \geq 0, \quad g_0(\theta) \geq R_0, \quad \beta_4\big(g_0(\theta) - R_0\big) = 0, \tag{17l}$$

$$\beta_5 \geq 0, \quad g_{02}(\theta) \geq R_0 + R_2, \quad \beta_5(g_{02}(\theta) - R_0 - R_2) = 0, \tag{17m}$$

$$\beta_6 \geq 0, \quad g_{012}(\theta) \geq R_0 + \sum_{k=1}^{2} R_k, \quad \beta_6\big(g_{012}(\theta) - R_0 - \sum_{k=1}^{2} R_k\big) = 0, \tag{17n}$$

$$\beta_7 \geq 0, \quad g_{0123}(\theta) \geq R_0 + \sum_{k=1}^{3} R_k, \quad \beta_7\big(g_{0123}(\theta) - R_0 - \sum_{k=1}^{3} R_k\big) = 0, \tag{17o}$$

$$\beta_8 \geq 0, \quad h_0(\theta) \geq R_0, \quad \beta_8(h_0(\theta) - R_0) = 0, \tag{17p}$$

$$\beta_9 \geq 0, \quad h_{03}(\theta) \geq R_0 + R_3, \quad \beta_9(g_{03}(\theta) - R_0 - R_3) = 0, \tag{17q}$$

$$\beta_{10} \geq 0, \quad h_{023}(\theta) \geq R_0 + \sum_{k=2}^{3} R_k, \quad \beta_{10}(h_{023}(\theta) - R_0 - \sum_{k=2}^{3} R_k) = 0, \tag{17r}$$

$$\beta_{11} \geq 0, \quad h_{0123}(\theta) \geq R_0 + \sum_{k=1}^{3} R_k, \quad \beta_{11}\big(h_{0123}(\theta) - R_0 - \sum_{k=1}^{3} R_k\big) = 0. \tag{17s}$$

The constraints on the left hand side of (17h)–(17s) are non-negativity constraints on the Lagrange multipliers, $\{\beta_i\}$. The middle constraints are the feasibility constraints of the rates and partitions and the constraints on the right hand side of (17h)–(17s) are the complementarity slackness conditions [27].

## V. An outer bound on $\text{Region}_1$

In this section we provide a relaxation of the optimization problem in (15). This relaxation will be shown to be tight in the sense described in Section VII.

Let us introduce three power partition vectors $\alpha, \alpha'$ and $\alpha''$. An outer bound on $\text{Region}_1$ can be obtained by solving the following optimization problem, in which $\alpha$ is employed in the constraints involving $f_0$, $f_{01}$, $f_{012}$ and $f_{0123}$, $\alpha'$ is employed in the constraints involving $g_0$, $g_{02}$, $g_{012}$ and $g_{0123}$, and $\alpha''$ is employed in the constraints involving $h_0$, $h_{03}$, $h_{023}$ and $h_{0123}$.

$$\max_{\{R_k\}_{k=1}^3, \alpha, \alpha', \alpha''} \quad \sum_{k=1}^{3} w_k R_k, \tag{18a}$$

$$\text{subject to} \quad R_0 \leq f_0(\alpha), \quad R_0 + R_1 \leq f_{01}(\alpha), \quad R_0 + R_1 + R_2 \leq f_{012}(\alpha), \tag{18b}$$

$$R_0 + R_1 + R_2 + R_3 \leq f_{0123}(\alpha), \tag{18c}$$

$$R_0 \leq g_0(\alpha'), \quad R_0 + R_2 \leq g_{02}(\alpha'), \quad R_0 + R_1 + R_2 \leq g_{012}(\alpha'), \tag{18d}$$

$$R_0 + R_1 + R_2 + R_3 \leq g_{0123}(\alpha'), \tag{18e}$$

$$R_0 \leq h_0(\alpha''), \quad R_0 + R_3 \leq h_{03}(\alpha''), \quad R_0 + R_2 + R_3 \leq h_{023}(\alpha''), \tag{18f}$$

$$R_0 + R_1 + R_2 + R_3 \leq h_{0123}(\alpha''), \tag{18g}$$

$$R_k \geq 0, k = 1, 2, 3, \tag{18h}$$

$$\alpha, \alpha', \alpha'' \in \bar{\mathcal{S}}. \tag{18i}$$

The optimization problem in (18) is a relaxation of that in (15) because (18) can be made equivalent to the problem in (15) by adding the constraint $\alpha = \alpha' = \alpha''$. Hence, for a given set of weights, the weighted sum-rate generated by (18) is greater than or equal to that generated by (15). However, in the following analysis we will show that for all relevant non-negative weights, $w_1$, $w_2$ and $w_3$, and allocated powers $P_1$ and $P_2$, any maximum weighted sum-rate generated by (18) is equal to the maximum weighted sum-rate generated by the problem in (15).

### A. *KKT conditions for* (18)

In order to gain insight into the structure of (18), we consider its KKT optimality conditions. Akin to Section IV-B, we will consider solutions to the KKT conditions for (18) for which for which $R > 0$, $(\alpha, \alpha', \alpha'') > 0$, $\sum_{\ell=1}^{2} \alpha_i^\ell < 1$, $\sum_{\ell=1}^{2} \alpha_i'^\ell < 1$ and $\sum_{\ell=1}^{2} \alpha_i''^\ell < 1$.

Using $L$ to denote the Lagrangian of (18), we can write the KKT conditions as follows:

$$\frac{\partial L}{\partial R_1} = w_1 - (\lambda_{01} + \lambda_{012} + \lambda_{0123}) - (\eta_{012} + \eta_{0123}) - \xi_{0123} = 0, \tag{19a}$$

$$\frac{\partial L}{\partial R_2} = w_2 - (\lambda_{012} + \lambda_{0123}) - (\eta_{02} + \eta_{012} + \eta_{0123}) - (\xi_{023} + \xi_{0123}) = 0, \tag{19b}$$

$$\frac{\partial L}{\partial R_3} = w_3 - \lambda_{0123} - \eta_{0123} - (\xi_{03} + \xi_{023} + \xi_{0123}) = 0, \tag{19c}$$

$$\frac{\partial L}{\partial \alpha_i^\ell} = \lambda_0 \frac{\partial f_0(\alpha)}{\partial \alpha_i^\ell} + \lambda_{01} \frac{\partial f_{01}(\alpha)}{\partial \alpha_i^\ell} + \lambda_{012} \frac{\partial f_{012}(\alpha)}{\partial \alpha_i^\ell} + \lambda_{0123} \frac{\partial f_{0123}(\alpha)}{\partial \alpha_i^\ell} + \mu_i = 0, \tag{19d}$$

$$\frac{\partial L}{\partial \alpha_i'^\ell} = \eta_0 \frac{\partial g_0(\alpha')}{\partial \alpha_i'^\ell} + \eta_{02} \frac{\partial g_{02}(\alpha')}{\partial \alpha_i'^\ell} + \eta_{012} \frac{\partial g_{012}(\alpha')}{\partial \alpha_i'^\ell} + \eta_{0123} \frac{\partial g_{0123}(\alpha')}{\partial \alpha_i'^\ell} + \mu_i' = 0, \tag{19e}$$

$$\frac{\partial L}{\partial \alpha_i''^\ell} = \xi_0 \frac{\partial h_0(\alpha'')}{\partial \alpha_i''^\ell} + \xi_{03} \frac{\partial h_{03}(\alpha'')}{\partial \alpha_i''^\ell} + \xi_{023} \frac{\partial h_{023}(\alpha'')}{\partial \alpha_i''^\ell} + \xi_{0123} \frac{\partial h_{0123}(\alpha'')}{\partial \alpha_i''^\ell} + \mu_i'' = 0, \tag{19f}$$

$$\lambda_0 \geq 0, \quad f_0(\alpha) \geq R_0, \quad \lambda_0(f_0(\alpha) - R_0) = 0, \tag{19g}$$

$$\lambda_{01} \geq 0, \quad f_{01}(\alpha) \geq R_0 + R_1, \quad \lambda_{01}(f_{01}(\alpha) - R_0 - R_1) = 0, \tag{19h}$$

$$\lambda_{012} \geq 0, \quad f_{012}(\alpha) \geq R_0 + R_1 + R_2, \quad \lambda_{012}(f_{012}(\alpha) - R_0 - R_1 - R_2) = 0, \tag{19i}$$

$$\lambda_{0123} \geq 0, \quad f_{0123}(\alpha) \geq R_0 + R_1 + R_2 + R_3, \quad \lambda_{0123}(f_{0123}(\alpha) - R_0 - R_1 - R_2 - R_3) = 0, \tag{19j}$$

$$\eta_0 \geq 0, \quad g_0(\alpha') \geq R_0, \quad \eta_0(g_0(\alpha') - R_0) = 0, \tag{19k}$$

$$\eta_{02} \geq 0, \quad g_{02}(\alpha') \geq R_0 + R_2, \quad \eta_{02}(g_{02}(\alpha') - R_0 - R_2) = 0, \tag{19l}$$

$$\eta_{012} \geq 0, \quad g_{012}(\alpha') \geq R_0 + R_1 + R_2, \quad \eta_{012}(g_{012}(\alpha') - R_0 - R_1 - R_2) = 0, \tag{19m}$$

$$\eta_{0123} \geq 0, \quad g_{0123}(\alpha') \geq R_0 + R_1 + R_2 + R_3, \quad \eta_{0123}(g_{0123}(\alpha') - R_0 - R_1 - R_2 - R_3) = 0, \tag{19n}$$

$$\xi_0 \geq 0, \quad h_0(\alpha'') \geq R_0, \quad \xi_0(h_0(\alpha'') - R_0) = 0, \tag{19o}$$

$$\xi_{03} \geq 0, \quad h_{03}(\alpha'') \geq R_0 + R_3, \quad \xi_{03}(h_{03}(\alpha'') - R_0 - R_3) = 0, \tag{19p}$$

$$\xi_{023} \geq 0, \quad h_{023}(\alpha'') \geq R_0 + R_2 + R_3, \quad \xi_{023}(h_{023}(\alpha'') - R_0 - R_2 - R_3) = 0, \tag{19q}$$

$$\xi_{0123} \geq 0, \quad h_{0123}(\alpha'') \geq R_0 + R_1 + R_2 + R_3, \quad \xi_{0123}(h_{0123}(\alpha'') - R_0 - R_1 - R_2 - R_3) = 0. \tag{19r}$$

In writing (19) we have used $\lambda_0, \lambda_{01}, \lambda_{012}, \lambda_{0123}$ to denote the Lagrange multipliers associated with the constraints involving $\alpha$ in (18), $\eta_0, \eta_{02}, \eta_{012}, \eta_{0123}$ to denote the multipliers associated with the constraints involving $\alpha'$, and $\xi_{03}, \xi_{023}, \xi_{0123}$ to denote the multipliers associated with the constraints involving $\alpha''$.

### *B. Analysis of the KKT system in* (19)

Because of the (partial) decoupling of constraints, it is significantly easier to draw insight into the system of equations in (19) than it is to draw insight into the KKT system for (15), cf. (17). In particular, we have the following results.

*Lemma 2:* Any solution of the KKT system in (19) must satisfy

$$\lambda_0 = \eta_0 = \xi_0 = 0. \tag{20}$$

*Proof:* See Appendix I-A. ∎

*Lemma 3:* For any solution of the KKT system in (19), either $\lambda_{01} = \lambda_{012} = \lambda_{0123} = 0$ or $\lambda_{01} > 0$, $\lambda_{012} > 0$ and $\lambda_{0123} > 0$.

*Proof:* See Appendix I-B. ∎

*Lemma 4:* For any solution of the KKT system in (19), either $\eta_{02} = \eta_{012} = \eta_{0123} = 0$ or $\eta_{02} > 0$ and $\eta_{0123} > 0$.

*Proof:* See Appendix I-C. ∎

*Lemma 5:* For any solution of the KKT system in (19), either $\xi_{03} = \xi_{023} = \xi_{0123} = 0$ or $\xi_{03} > 0$, $\xi_{023}$ and $\xi_{0123} > 0$.

*Proof:* See Appendix I-D. ∎

## VI. Common solutions for KKT conditions in (17) and (19)

In this section we use the results in Lemmas 2–5 to construct an explicit characterization of solutions of the KKT system in (17) that also solve the KKT system in (19). In particular, we have:

*Theorem 1 ($w_1 > w_2 > w_3$):* Given $R_0$ satisfying (14) and a weight vector with $w_1 > w_2 > w_3$, there exists a solution of the KKT system of equations in (17), $(\beta, \theta, R)$, with $\theta > 0$ and $R > 0$ such that this solution solves the KKT system in (19) with $(\lambda, \eta, \xi) = \beta$, $\alpha = \theta$ and with identical rates $R$.

*Proof:* See Appendix II. ∎

*Theorem 2 ($w_2 > w_1 > w_3$):* Given $R_0$ satisfying (14) and a weight vector with $w_2 > w_1 > w_3$, there exists a solution of the KKT system in (17), $(\beta, \theta, R)$, with $\theta > 0$ and $R > 0$ such that this solution solves the KKT system in (19) with $(\lambda, \eta, \xi) = \beta$, $\alpha' = \theta$ and with identical rates $R$.

*Proof:* See Appendix IV. ∎

*Theorem 3 ($w_1 > w_3 > w_2$):* Given $R_0$ satisfying (14) and a weight vector with $w_1 > w_3 > w_2$, any locally optimal solution of (15) must have $R_2 = 0$. Furthermore, for this weight ordering, the optimal solution of (18) must have $R_2 = 0$.

*Proof:* See Appendix V. ■

*Corollary 1 ($w_1 > w_3 > w_2$):* Given $R_0$ satisfying (14) and a weight vector with $w_1 > w_3 > w_2$, there exists a solution of the KKT system in (17), $(\beta, \theta, R)$, with $\theta_1^2 = \theta_2^2 = 0$ and $R_2 = 0$ such that this solution solves the KKT system in (19) with $(\lambda, \eta, \xi) = \beta$, $\alpha' = \theta$ and with identical rates $R$.

*Proof:* First observe that with $R_2 = 0$, the partitions $\theta_1^2$ and $\theta_2^2$ can be set to zero without affecting the set of rates generated by (15). To prove this corollary, we consider the KKT conditions corresponding to (15) with $\theta_1^2$, $\theta_2^2$ and $R_2$ set to zero. Doing so, it is straightforward to show that one solution of the resulting KKT conditions has $\beta_i = 0$, $i = 0, \ldots, 4, 8, 9, 11$. (In this case the KKT conditions do not involve $\beta_2, \beta_5$ and $\beta_{10}$.) The statement of the corollary follows from identifying $(\lambda, \eta, \xi)$ with $\beta$, $\alpha'$ with $\theta$ and noting that (18) yields the same rates as (15). ■

*Remark 2 (Other cases):*

- When $w_3 > w_2 > w_1$, the conditions in (19a)–(19c) yield:

$$\beta_9 > \beta_2 + \beta_5 + \beta_6 \geq 0 \qquad \text{and} \qquad \beta_5 + \beta_{10} > \beta_1 \geq 0.$$

  Using these inequalities and the symmetry between receivers $Y$ and $W$, a result analogous to the one in Theorem 1 can be derived *mutatis mutandis*.
- When $w_3 > w_1 > w_2$, the methodology used to arrive at Theorem 3 and Corollary 1 can be used to show that any locally optimal solution of (15) must have $R_2 = 0$, and that such a solution solves the KKT system in (19).
- When $w_2 > w_3 > w_1$, we show in Appendix VI that there is no solution of the system of equations in (17) that lies in the feasible set of (16), i.e., no solution that lies in Region$_2$. Therefore, for this weight ordering the rates and partitions generated by solving (17) (corresponding to (15)) do not lie in the SPC region (cf. (13)), and we do not need to study this case further. From the symmetry between receivers $Y$ and $W$, one can see that for this weight ordering the SPC rates and partitions are generated by solving the KKT conditions corresponding to (16) in a way analogous to that used for Region$_1$ when $w_2 > w_1 > w_3$. See Remark 3, below. □

An observation regarding Theorems 1–3, Corollary 1 and Remark 2 is that for every weight ordering only one of the partitions $\alpha$, $\alpha'$ or $\alpha''$ is used to yield a rate vector on the boundary of the SPC rate region.

## VII. Tightness of the relaxation in (18)

In this section we show that for all relevant weight orderings, i.e., all orderings except $w_2 > w_3 > w_1$, the rate region generated by (18) is identical to that generated by (15). To show that, we have the following result, which is based on a transformation of the problem in (18) into a convex optimization problem.

*Theorem 4:* The KKT conditions are necessary and sufficient optimality conditions for the problem in (18).

*Proof:* See Appendix VIII. ■

Theorem 4 implies that for any $R_0$ satisfying (14) and any relevant weight vector, $w$, any solution of the KKT system in (19) yields the maximum weighted sum-rate. This leads to

*Theorem 5:* The rate region generated by solving (18) for all relevant weight vectors, i.e., all orderings except $w_2 > w_3 > w_1$, is identical to that generated by solving (15).

*Proof:* Recall that the solutions provided in Theorems 1, 2, 3 and Corollary 1 solve both the KKT conditions corresponding to (15) and those corresponding to (18). Using Theorem 4, we conclude that these solutions are sufficient for the optimality of the weighted sum-rate generated in (18). Now, (18) is a relaxation of (15), and hence the weighted sum-rate generated by (18) is an upper bound on that generated by (15). Since the solutions provided Theorems 1, 2, 3 and Corollary 1 yield the identical rates for both (15) and (18), we conclude that these solutions yield the maximum weighted sum-rate in (15) and hence the theorem. ■

So far we have shown that for any distinct weight settings, apart from the case in which $w_2 > w_3 > w_1$, the optimization problems in (15) and (18) are equivalent under the assumption that the power partitions are strictly greater than zero. This, however, is not restrictive, because the rate constraints are continuous functions of the power partitions. Hence, infinitesimal changes in the power partitions result in infinitesimal changes in the data rates.

Using the symmetry between receivers $Y$ and $W$, it follows that

*Theorem 6:* For all relevant weight vectors, i.e., all orderings except $w_2 > w_1 > w_3$, the rate region generated by solving (16) is identical to that generated by solving the corresponding relaxation.

*Proof:* The proof of this theorem follows a path analogous to that used to prove Theorem 5. ■

*Remark 3:* For all weight orderings other than $w_2 > w_3 > w_1$, solving (18) yields SPC achievable rates. When $w_2 > w_3 > w_1$, SPC achievable rates can be obtained by solving the corresponding relaxation of Region$_2$. Alternatively, for all weight orderings other than $w_2 > w_1 > w_3$ SPC achievable rates can be obtained by solving the corresponding relaxation of Region$_2$, and when $w_2 > w_1 > w_3$, SPC achievable rates can be obtained by solving (18). □

Our goal now is to show that for any achievable rate vector $R$, there exists a power partition $\theta \in \mathcal{S}$ such that $(\theta, R) \in \mathrm{Region}_i$, $i = 1, 2$. The equivalence between (15) and (18) implies that (for Region$_1$) it is sufficient to show that for any achievable rate vector $R$, there exists power partitions $\alpha, \alpha', \alpha'' \in \mathcal{S}$, such that the constraints (18b)–(18i) are satisfied. An analogous argument can be used to show that Region$_2$ contains all achievable partition-rate vectors.

## VIII. Information theoretic bounds on achievable rates

In this section we provide some information theoretic bounds on the achievable rates. These bounds will be used in Section IX to show that Region$_1$ and Region$_2$ contain all achievable rate vectors.

Let $M_0 \in \{1, \ldots, 2^{nR_0}\}$ denote the common message that is intended for all receivers, and let $M_k \in \{1, \ldots, 2^{nR_k}\}$, $k = 1, 2, 3$ denote the particular messages of receivers $Y$, $Z$ and $W$, respectively; see Figure 1. Let the decoder of receiver $Y$ be denoted by $g_1$, where $g_1$ maps a length $n$ block of the signal received by receiver $Y$ to the set of receiver $Y$'s messages; that is, $g_1 : (Y_1, Y_2) \mapsto (\hat{M}_0, \hat{M}_1)$. An error event for receiver $Y$ occurs if $(\hat{M}_0, \hat{M}_1) \neq (M_0, M_1)$. The average probability of this event is denoted by $P_{e_1}^n$. In a similar manner, we define the decoders and the associated average error probabilities of receivers $Z$ and $W$ and denote them by $P_{e_2}^n$ and $P_{e_3}^n$, respectively. A rate vector $(R_0, R_1, R_2, R_3)$ is achievable if for every $\epsilon > 0$ there exists a sequence of codes (indexed by $n$) such that for all sufficiently large $n$ the probability of error $P_e^n < \epsilon$, where $P_e^n = \max\{P_{e_1}^n, P_{e_2}^n, P_{e_3}^n\}$.

To obtain the desired information theoretic bounds on achievable rate vectors, let $\epsilon_i$, $i = 1, 2, 3$ be a small positive number and let

$$\begin{aligned}
&\mathcal{U}_1^3 = [M_0, M_3, Z_2, Y_2], && \mathcal{V}_1^3 = [M_0, Z_2], && \mathcal{X}_1^3 = \mathcal{V}_1^3, \\
&\mathcal{U}_2^3 = [M_0, M_1, Z_1, W_1], && \mathcal{V}_2^3 = [M_0, Z_1], && \mathcal{X}_2^3 = \mathcal{V}_2^3, \\
&\mathcal{U}_1^2 = [\mathcal{U}_1^3, M_2], && \mathcal{V}_1^2 = [\mathcal{V}_1^3, M_2], && \mathcal{X}_1^2 = [\mathcal{X}_1^3, M_2, M_3, W_2], \\
&\mathcal{U}_2^2 = [\mathcal{U}_2^3, M_2], && \mathcal{V}_2^2 = [\mathcal{V}_2^3, M_2, M_1, Y_1], && \mathcal{X}_2^2 = [\mathcal{X}_2^3, M_2].
\end{aligned} \tag{21}$$

In Appendix IX we will use Fano's inequality to show that

$$nR_0 \leq I(\mathcal{U}_1^3; Y_1) + I(\mathcal{U}_2^3; Y_2) + n\epsilon_1, \tag{22a}$$

$$n(R_0 + R_1) \leq I(\mathcal{U}_2^3; Y_2) + I(X_1; Y_1) + n\epsilon_1, \tag{22b}$$

$$n(R_0 + R_1 + R_2) \leq I(\mathcal{U}_2^3; Y_2) + I(\mathcal{U}_2^2; Z_2 | \mathcal{U}_2^3) + I(X_1; Y_1) + n\epsilon_1 + n\epsilon_2, \tag{22c}$$

$$\begin{aligned}
n(R_0 + R_1 + R_2 + R_3) \leq\ & I(\mathcal{U}_2^3; Y_2) + I(\mathcal{U}_2^2; Z_2 | \mathcal{U}_2^3) + I(X_2; W_2 | \mathcal{U}_2^2) \\
& + I(X_1; Y_1) + n\epsilon_1 + n\epsilon_2 + n\epsilon_3,
\end{aligned} \tag{22d}$$

$$nR_0 \leq I(\mathcal{U}_1^3; W_1) + I(\mathcal{U}_2^3; W_2) + n\epsilon_3, \tag{22e}$$

$$n(R_0 + R_3) \leq I(\mathcal{U}_1^3; W_1) + I(X_2; W_2) + n\epsilon_3, \tag{22f}$$

$$n(R_0 + R_2 + R_3) \leq I(\mathcal{U}_1^3; W_1) + I(\mathcal{U}_1^2; Z_1|\mathcal{U}_1^3) + I(X_2; W_2) + n\epsilon_2 + n\epsilon_3, \tag{22g}$$

$$\begin{aligned} n(R_0 + R_1 + R_2 + R_3) \leq\ & I(\mathcal{U}_1^3; W_1) + I(\mathcal{U}_1^2; Z_1|\mathcal{U}_1^3) + I(X_1; Y_1|\mathcal{U}_1^2) \\ & + I(X_2; W_2) + n\epsilon_1 + n\epsilon_2 + n\epsilon_3, \end{aligned} \tag{22h}$$

where $X_i$ is the symbol transmitted on subchannel $i$; cf. Figure 1.

In Appendix X we show that

$$nR_0 \leq I(\mathcal{V}_1^3; Z_1) + I(\mathcal{V}_2^3; Z_2) + n\epsilon_2, \tag{23a}$$

$$n(R_0 + R_2) \leq I(\mathcal{V}_1^2; Z_1) + I(\mathcal{V}_2^2; Z_2) + n\epsilon_2, \tag{23b}$$

$$n(R_0 + R_1 + R_2) \leq I(X_1; Y_1|\mathcal{V}_1^2) + I(\mathcal{V}_1^2; Z_1) + I(\mathcal{V}_2^2; Z_2) + n\epsilon_1 + n\epsilon_2, \tag{23c}$$

$$\begin{aligned} n(R_0 + R_1 + R_2 + R_3) \leq\ & I(X_1; Y_1|\mathcal{V}_1^2) + I(X_2; W_2|\mathcal{V}_2^2) + I(\mathcal{V}_1^2; Z_1) \\ & + I(\mathcal{V}_2^2; Z_2) + n\epsilon_1 + n\epsilon_2 + n\epsilon_3. \end{aligned} \tag{23d}$$

In Appendix X we also show that

$$nR_0 \leq I(\mathcal{X}_1^3; Z_1) + I(\mathcal{X}_2^3; Z_2) + n\epsilon_2, \tag{24a}$$

$$n(R_0 + R_2) \leq I(\mathcal{X}_1^2; Z_1) + I(\mathcal{X}_2^2; Z_2) + n\epsilon_2, \tag{24b}$$

$$n(R_0 + R_2 + R_3) \leq I(X_2; W_2|\mathcal{X}_2^2) + I(\mathcal{X}_1^2; Z_1) + I(\mathcal{X}_2^2; Z_2) + n\epsilon_1 + n\epsilon_2, \tag{24c}$$

$$\begin{aligned} n(R_0 + R_1 + R_2 + R_3) \leq\ & I(X_1; Y_1|\mathcal{X}_1^2) + I(X_2; W_2|\mathcal{X}_2^2) + I(\mathcal{X}_1^2; Z_1) \\ & + I(\mathcal{X}_2^2; Z_2) + n\epsilon_1 + n\epsilon_2 + n\epsilon_3. \end{aligned} \tag{24d}$$

The inequalities in (22a)–(22d) and (23) will be used to show that Region$_1$ is an outer bound on the capacity region, whereas for Region$_2$, we will use (22a)–(22d) and (24).

## IX. The capacity region of the BC channel in Figure 1

From Section VII we have that the rate regions generated by Region$_1$ and Region$_2$ are equivalent to those generated by their corresponding relaxations. Hence, to show the converse it suffices to show that the rate vectors contained in the intersection of these relaxations form an outer bound on the capacity region. That is, for every achievable rate vector there exist three independent sets of power partitions such that the inequalities in (18b)–(18i), and those corresponding to Region$_2$ are satisfied. In order to do that we invoke the inequalities in (22) and (23) in the case in which the subchannels are Gaussian. Using (22) and (24), an analogous argument can be used to show that $\text{Region}_2$ is an outer bound on the capacity region. For $\text{Region}_1$ we have

*Theorem 7:* The rate region generated by (18) is an outer bound on the achievable rate region.

*Proof:* See Appendix XI. ■

For $\text{Region}_2$ we have

*Theorem 8:* The rate region generated by the corresponding relaxation of $\text{Region}_2$ in (16) is an outer bound on the achievable rate region.

*Proof:* The proof of this theorem parallels that of Theorem 7 but with $\alpha'$ defined as in Appendix XII. ■

We are now ready to present the main result of the paper.

*Theorem 9:* The capacity region of the BC channel in Figure 1 is the closure of the region of rates achieved by superposition coding and Gaussian signalling, which is the closure of rate vectors contained in the SPC region defined in (2).

*Proof:*

- Achievability: By construction, for any given power allocation, all rate vectors satisfying (2) are achievable by superposition coding and Gaussian signalling.
- Converse: To prove that the SPC region contains all achievable rates, we observe that the SPC region in (2) is equivalent to $\text{Region}_1 \bigcap \text{Region}_2$; cf. (13). From Theorem 5 and Theorem 6, we have that the rate vectors in $\text{Region}_1$ and $\text{Region}_2$ are identical to those generated by the corresponding relaxations. Now, from Theorem 7, we have that the rate region generated by (18) is an outer bound on the achievable rate region. Using a similar argument with Theorem 8, we see that the rate vectors contained in $\text{Region}_2$ also form an outer bound on the achievable rate region. Hence, the rate vectors contained in $\text{Region}_1 \bigcap \text{Region}_2$ form an outer bound on the achievable rate region, which establishes the desired converse.

■

## X. Conclusion

This paper considered the class of BC channels depicted in Figure 1, wherein each receiver receives a particular message along with a common message that is intended to all receivers. It was shown that, for this scenario, every achievable rate vector can be attained by superposition coding and Gaussian signalling. Our approach to establishing this result is based on an ostensibly relaxed characterization of the region of rates that can be achieved by superposition coding and Gaussian signalling and on showing that this relaxation is tight. Although the focus of this work has been restricted to the scenario depicted in Figure 1, we suspect that the same methodology can be applied to systems with alternate degradation orders and possibly with more receivers.

## APPENDIX I

## ANALYSIS OF THE LAGRANGE MULTIPLIERS IN (19)

### A. *Proof of Lemma 2*

In order to find $\lambda_0$, we use the fact that $f_{01}(\alpha)$, $f_{012}(\alpha)$ and $f_{0123}(\alpha)$ are not functions of $\alpha_1^\ell$, and hence from (19d) we have

$$\frac{\partial L}{\partial \alpha_1^\ell} = \frac{-\lambda_0 P_1}{N_1^1 + (\alpha_1^1 + \alpha_1^2)P_1} = 0. \tag{25}$$

Hence, for any $P_1 > 0$, $\lambda_0 = 0$. For $\eta_0$, we apply the observation that $g_{02}(\alpha'), g_{012}(\alpha')$ and $g_{0123}(\alpha')$ are not functions of $\alpha_i'^2$ to (19e) to write

$$\frac{\partial L}{\partial \alpha_1'^2} = \frac{-\eta_0 P_1}{N_1^2 + (\alpha_1'^2 + \alpha_1'^1)P_1} = 0, \tag{26}$$

which yields $\eta_0 = 0$ for any $P_1 > 0$. Similarly, by differentiating the Lagrangian with respect to $\alpha_2''^\ell$, and using (19f), one can show that for any $P_2 > 0$, $\xi_0 = 0$.

### B. *Proof of Lemma 3:*

In order to draw some insight into the relationship between these multipliers, we use (19d) and the defintions in Section II to write

$$\frac{\partial L}{\partial \alpha_2^1} = \frac{-(\lambda_{01} + \lambda_{012} + \lambda_{0123})P_2}{N_2^3 + (\alpha_2^1 + \alpha_2^2)P_2} + \frac{-(\lambda_{012} + \lambda_{0123})P_2 \alpha_2^2}{(N_2^2 + (\alpha_2^1 + \alpha_2^2)P_2)(N_2^2 + \alpha_2^1 P_2)} + \frac{\lambda_{0123} P_2}{N_2^1 + \alpha_2^1 P_2} = 0, \tag{27}$$

$$\frac{\partial L}{\partial \alpha_2^2} = \frac{-(\lambda_{01} + \lambda_{012} + \lambda_{0123})P_2}{N_2^3 + (\alpha_2^1 + \alpha_2^2)P_2} + \frac{(\lambda_{012} + \lambda_{0123})P_2}{N_2^2 + (\alpha_2^1 + \alpha_2^2)P_2} = 0. \tag{28}$$

Since, $P_2 > 0$ and $\lambda_{01}, \lambda_{012}$ and $\lambda_{0123}$ are non-negative, (27) and (28) are satisfied if and only if $\lambda_{01} = \lambda_{012} = \lambda_{0123} = 0$ or

$$\lambda_{0123} > 0 \qquad \text{and} \qquad \lambda_{01} > 0. \tag{29}$$

Furthermore, substituting from (28) into (27), we have

$$\frac{\lambda_{012} + \lambda_{0123}}{N_2^2 + \alpha_2^1 P_2} = \frac{\lambda_{0123}}{N_2^1 + \alpha_2^1 P_2}. \tag{30}$$

Since $N_2^2 > N_2^1$, this implies that

$$\lambda_{012} > 0, \tag{31}$$

unless $\lambda_{01} = \lambda_{012} = \lambda_{0123} = 0$.

*C. Proof of Lemma 4*

Using (19e) we have

$$\frac{\partial L}{\partial \alpha_1'^1} = \frac{-(\eta_{02}+\eta_{012}+\eta_{0123})P_1}{N_1^2+\alpha_1'^1 P_1} + \frac{(\eta_{012}+\eta_{0123})P_1}{N_1^1+\alpha_1'^1 P_1} = 0, \tag{32}$$

$$\frac{\partial L}{\partial \alpha_2'^1} = \frac{-(\eta_{02}+\eta_{012}+\eta_{0123})P_2}{N_2^2+\alpha_2'^1 P_2} + \frac{\eta_{0123}P_2}{N_2^1+\alpha_2'^1 P_2} = 0. \tag{33}$$

Since $P_1$ and $P_2$ are strictly greater than zero, from (32) and (33) and the non-negativity of $\eta_{02}, \eta_{012}$ and $\eta_{0123}$, we have either $\eta_{02} = \eta_{012} = \eta_{0123} = 0$, or

$$\eta_{02} > 0 \qquad \text{and} \qquad \eta_{0123} > 0. \tag{34}$$

*D. Proof of Lemma 5*

Using the definitions in Section II, we differentiate the Lagrangian with respect to $\alpha_1''^1$ and $\alpha_1''^2$. Substituting into (19f), we conclude that either $\xi_{02} = \xi_{023} = \xi_{0123} = 0$, or

$$\xi_{03} > 0, \qquad \xi_{023} > 0 \qquad \text{and} \qquad \xi_{0123} > 0. \tag{35}$$

## Appendix II
## Proof of Theorem 1

*A. A solution of KKT system of equations* (17) *for original problem* (15) *for* $w_1 > w_2 > w_3$

For $w_1 > w_2 > w_3$, (17a)–(17c) yield

$$\beta_1 > \beta_5 + \beta_{10} \geq 0, \quad \text{and} \quad \beta_2 + \beta_5 + \beta_6 > \beta_9 \geq 0. \tag{36}$$

From the first inequality in (36), it is seen that $\beta_1 > 0$. Using this fact in (17e), it is seen that $\beta_2 + \beta_3 > 0$. For the moment, we will assume that $\beta_2 > 0$ and $\beta_3 > 0$ and we will show later that this assumption is without loss of generality.

In Appendix III we show that one solution of the KKT conditions can be obtained by setting

$$\beta_i = 0, \; i = 0, 4, \ldots, 11. \tag{37}$$

Using (37) we have $w_1 = \beta_1 + \beta_2 + \beta_3$, $w_2 = \beta_2 + \beta_3$, and $w_3 = \beta_3$. The complementarity slackness conditions for this choice of Lagrange multipliers yield

$$R_0 + R_1 = f_{01}(\theta), \quad R_0 + R_1 + R_2 = f_{012}(\theta), \quad \text{and} \quad R_0 + R_1 + R_2 + R_3 = f_{0123}(\theta). \tag{38}$$

We now show that because $\beta_2 + \beta_3 > 0$, we can assume that $\beta_2 > 0$ and $\beta_3 > 0$, without loss of generality. Towards that end, we observe that:

i. If $\beta_2 > 0$ and $\beta_3 = 0$, then $R_2$ is determined by the second equality in (38). Now, in this case, it may not be immediately clear that $R_3$ is determined by the last equality in (38). Substituting from (17e) into (17d) we have that $\beta_7 > 0$, which implies that $R_0 + R_1 + R_2 + R_3 = g_{0123}(\theta)$.

Now, suppose that $R_3$ is not determined by the last equality in (38). In that case we would have $R_3 < \frac{1}{2}\log\Big(\frac{N_2^1+\theta_2^1 P_2}{N_2^1}\Big)$. Since $R_0 + R_1 + R_2 + R_3 = g_{0123}(\theta)$, it would follow that $R_0 + R_1 + R_2 > g_{012}(\theta)$, which contradicts the middle inequality in (17n). Hence, in this case $R_3$ must be determined by the equalities in (38).

ii. If $\beta_2 = 0$ and $\beta_3 > 0$, then $R_2 + R_3$ is determined by the third equality in (38). Furthermore, from the last two terms in (17d) we have that $\beta_5 + \beta_6 > 0$. We will use contradiction to show that $\beta_5 = \beta_6 = 0$.

Suppose that $\beta_5 > 0$. In that case, $R_0 + R_2 = g_{02}(\theta)$. From the middle inequality in (17j), we have $R_2 \le \frac{1}{2}\log\Big(\frac{N_2^2+(\theta_2^1+\theta_2^2)P_2}{N_2^2+\theta_2^1 P_2}\Big)$, because $\beta_1 > 0$ leads to $R_1$ being determined by the first equality in (38). Now,

$$\begin{aligned}
R_0 &\ge g_{02}(\theta) - \frac{1}{2}\log\Big(\frac{N_2^2+(\theta_2^1+\theta_2^2)P_2}{N_2^2+\theta_2^1 P_2}\Big)\\
&= \frac{1}{2}\log\Big(\frac{N_1^2+P_1}{N_1^2+\theta_1^1 P_1}\Big) + \frac{1}{2}\log\Big(\frac{N_2^2+P_2}{N_2^2+\theta_2^1 P_2}\Big) - \frac{1}{2}\log\Big(\frac{N_2^2+(\theta_2^1+\theta_2^2)P_2}{N_2^2+\theta_2^1 P_2}\Big)\\
&= \frac{1}{2}\log\Big(\frac{N_1^2+P_1}{N_1^2+\theta_1^1 P_1}\Big) + \frac{1}{2}\log\Big(\frac{N_2^2+P_2}{N_2^2+(\theta_2^1+\theta_2^2)P_2}\Big)\\
&> g_0(\theta),
\end{aligned}$$

which contradicts the middle inequality in (17l). Hence, we conclude that $\beta_5 = 0$.

Now, assume that $\beta_6 > 0$ and consider the last two terms in (17g). Since $\beta_5 = 0$, it is seen that $\beta_{10} > 0$. In that case the complementarity slackness condition in (17r) implies that $R_0 + R_2 + R_3 = h_{023}(\theta)$, but since $\beta_1 > 0$ and $\beta_3 > 0$, we have $R_2 + R_3 = \frac{1}{2}\log\Big(\frac{N_2^2+(\theta_2^1+\theta_2^2)P_2}{N_2^2+\theta_2^1 P_2}\Big) + \frac{1}{2}\log\Big(\frac{N_2^1+\theta_2^1 P_2}{N_2^1}\Big)$. Hence, in this case we have

$$\begin{aligned}
R_0 &= \frac{1}{2}\log\Big(\frac{N_1^3+P_1}{N_1^3+(\theta_1^1+\theta_1^2)P_1}\Big) + \frac{1}{2}\log\Big(\frac{N_2^1+P_2}{N_2^1}\Big) + \frac{1}{2}\log\Big(\frac{N_1^2+(\theta_1^1+\theta_1^2)P_1}{N_1^2+\theta_1^1 P_1}\Big)\\
&\quad - \frac{1}{2}\log\Big(\frac{N_2^2+(\theta_2^1+\theta_2^2)P_2}{N_2^2+\theta_2^1 P_2}\Big) - \frac{1}{2}\log\Big(\frac{N_2^1+\theta_2^1 P_2}{N_2^1}\Big)\\
&= \frac{1}{2}\log\Big(\frac{N_1^3+P_1}{N_1^3+(\theta_1^1+\theta_1^2)P_1}\Big) + \frac{1}{2}\log\Big(\frac{N_2^1+P_2}{N_2^1+\theta_2^1 P_2}\Big) + \frac{1}{2}\log\Big(\frac{N_1^2+(\theta_1^1+\theta_1^2)P_1}{N_1^2+\theta_1^1 P_1}\Big)\\
&\quad - \frac{1}{2}\log\Big(\frac{N_2^2+(\theta_2^1+\theta_2^2)P_2}{N_2^2+\theta_2^1 P_2}\Big)
\end{aligned}$$

$$
\begin{aligned}
&> \frac{1}{2}\log\Big(\frac{N_1^3+P_1}{N_1^3+(\theta_1^1+\theta_1^2)P_1}\Big)+\frac{1}{2}\log\Big(\frac{N_1^2+(\theta_1^1+\theta_1^2)P_1}{N_1^2+\theta_1^1P_1}\Big)+\frac{1}{2}\log\Big(\frac{N_2^1+P_2}{N_2^1+\theta_2^1P_2}\Big)\\
&\quad-\frac{1}{2}\log\Big(\frac{N_2^1+(\theta_2^1+\theta_2^2)P_2}{N_2^1+\theta_2^1P_2}\Big)\\
&=\frac{1}{2}\log\Big(\frac{N_1^3+P_1}{N_1^3+(\theta_1^1+\theta_1^2)P_1}\Big)+\frac{1}{2}\log\Big(\frac{N_1^2+(\theta_1^1+\theta_1^2)P_1}{N_1^2+\theta_1^1P_1}\Big)+\frac{1}{2}\log\Big(\frac{N_2^1+P_2}{N_2^1+(\theta_2^1+\theta_2^2)P_2}\Big)\\
&> h_0(\theta),
\end{aligned}
$$

which violates the middle inequality in (17p). Hence, we conclude that $\beta_6=0$, $\beta_2>0$, $\beta_3>0$, and that the rates $R_1$, $R_2$ and $R_3$ are determined by (38).

### B. *A solution of KKT system of equations* (19) *for relaxed problem* (18) *for* $w_1>w_2>w_3$

Using (19a) and (19b), we have

$$\lambda_{01}>\eta_{02}+\xi_{023}\geq 0, \tag{39}$$

and from (19b) and (19c)

$$\lambda_{012}+\eta_{02}+\eta_{012}>\xi_{03}\geq 0. \tag{40}$$

From (39), we have $\lambda_{01}>0$. Hence, from Lemma 3, we have

$$\lambda_{012}>0 \qquad \text{and} \qquad \lambda_{0123}>0. \tag{41}$$

Using (19h)–(19j), we have

$$R_0+R_1=f_{01}(\alpha),\quad R_0+R_1+R_2=f_{012}(\alpha)\quad \text{and}\quad R_0+R_1+R_2+R_3=f_{0123}(\alpha). \tag{42}$$

Now, we set

$$\eta_{02}=\eta_{012}=\eta_{0123}=\gamma_{02}=\gamma_{023}=\gamma_{0123}=\xi_{03}=\xi_{023}=\xi_{0123}=0. \tag{43}$$

Using (42) it can be seen that the setting in (43) solves the KKT system of equations in (19), for any $\alpha'$, and $\alpha''$ that satisfy the middle inequalities in (19k)–(19r).

### C. *Identifying solutions of* (17) *and* (19) *for* $w_1>w_2>w_3$

For $w_1>w_2>w_3$, we now compare the solution of the KKT system of the relaxed problem in the previous section with that of the KKT system of the original problem (15) in Appendix II-A. In particular, let

$$\lambda_0=\beta_0,\quad \lambda_{01}=\beta_1,\quad \lambda_{012}=\beta_2,\quad \lambda_{0123}=\beta_3, \tag{44}$$

$$\eta_0=\beta_4,\quad \eta_{02}=\beta_5,\quad \eta_{012}=\beta_6,\quad \eta_{0123}=\beta_7, \tag{45}$$

and

$$\xi_0 = \beta_8, \quad \xi_{02} = \beta_9, \quad \xi_{012} = \beta_{10}, \quad \xi_{0123} = \beta_{11}. \tag{46}$$

Since a solution of (17) exists with $\beta_i = 0,\ i = 0, 4, \ldots, 11$, it is seen that for this solution (17d) and (17e) become identical to (27) and (28), respectively, when

$$\alpha = \theta. \tag{47}$$

Now, we choose $\alpha'$ and $\alpha''$ arbitrarily so that (19k)–(19r) are satisfied. Such $\alpha'$ and $\alpha''$ exist because there exists a power partition $\theta$ that satisfies (17). Finally, it is seen that with $\alpha$ identified with $\theta$, the rates generated by (42) are identical to those generated by (38), which completes the proof of Theorem 1.

## APPENDIX III
## PROVING THAT FOR $w_1 > w_2 > w_3$, $\beta_i = 0,\ i = 0, 4, \ldots, 11$

When $w_1 > w_2 > w_3$, we have $\beta_1 > 0$, $\beta_2 > 0$ and $\beta_3 > 0$; see Appendix II-A. First, let us assume that $\beta_{10} + \beta_{11} > 0$. We will use contradiction to show that $\beta_{10} = \beta_{11} = 0$.

For this weight ordering $\beta_1 > 0$, $\beta_2 > 0$ and $\beta_3 > 0$, and hence the rates are determined by (17i)–(17k). Therefore, from the last equality in (17i) we have

$$R_1 = \frac{1}{2}\log\Big(\frac{N_1^1 + P_1}{N_1^1}\Big) + \frac{1}{2}\log\Big(\frac{N_2^3 + P_2}{N_2^3 + (\alpha_2^1 + \alpha_2^2)P_2}\Big) - R_0. \tag{48}$$

Let us assume that $\beta_{10}$ and $\beta_{11}$ can be arbitrarily small but $\beta_{10} > 0$ and $\beta_{11} > 0$. In that case, using (17r) and (17s) we have,

$$R_1 = \frac{1}{2}\log\Big(\frac{N_1^1 + \theta_1^1 P_1}{N_1^1}\Big). \tag{49}$$

Substituting from (49) into (48) yields

$$\begin{aligned} R_0 &= \frac{1}{2}\log\Big(\frac{N_2^3 + P_2}{N_2^3 + (\theta_2^1 + \theta_2^2)P_2}\Big) + \frac{1}{2}\log\Big(\frac{N_1^1 + P_1}{N_1^1}\Big) - \frac{1}{2}\log\Big(\frac{N_1^1 + \theta_1^1 P_1}{N_1^1}\Big) \\ &> \frac{1}{2}\log\Big(\frac{N_2^3 + P_2}{N_2^3 + (\theta_2^1 + \theta_2^2)P_2}\Big) + \frac{1}{2}\log\Big(\frac{N_1^1 + P_1}{N_1^1 + (\theta_1^1 + \theta_1^2)P_1}\Big) = f_0(\theta), \end{aligned}$$

which contradicts the middle constraint in (17h). Hence, it is seen that the case of $\beta_{10} > 0$ and $\beta_{11} > 0$ can be eliminated.

A similar argument can be used to eliminate the possibility that either $\beta_{10}$ or $\beta_{11}$ is greater than zero. If $\beta_{10} = 0$ and $\beta_{11} > 0$, then from (17g), $\beta_5 > 0$. Using the complementarity slackness condition associated with $\beta_5$ and the fact that $R_2$ is determined by the last equality in (17j), one can show that $R_0$ violates the middle constraint in (17l). Hence, the possibility of $\beta_{10} = 0$ and $\beta_{11} > 0$ can also be eliminated.

Now, let us consider the possibility that $\beta_{10} > 0$ and $\beta_{11} = 0$. In this case, we can assume that $R_1 \leq \frac{1}{2}\log\Big(\frac{N_1^1+\theta_1^1 P_1}{N_1^1}\Big)$. Using this expression in the first equality in (38) yields an $R_0$ that violates the middle inequality in (17h). Hence, the possibility that $\beta_{10} > 0$ and $\beta_{11} = 0$ can be eliminated.

In the argument above we have shown that $\beta_{10} + \beta_{11} = 0$. Since the last term on the left hand side of (17f) is the only non-negative term, we conclude that $\beta_0 = \beta_4 = \beta_8 = \beta_9 = 0$. Substituting $\beta_i = 0$ for $i = 0, 4, 8, 9, 10, 11$ into (17g) we have,

$$\frac{\beta_6 + \beta_7}{N_1^1 + \theta_1^1 P_1} = \frac{\beta_5 + \beta_6 + \beta_7}{N_1^2 + \theta_1^1 P_1}. \tag{50}$$

Now, we will use contradiction to show that $\beta_5 = 0$. To do that, we note that if $\beta_5 > 0$ then from the last equality in (17m)

$$R_0 + R_2 = \frac{1}{2}\log\Big(\frac{N_1^2 + P_1}{N_1^2 + \theta_1^1 P_1}\Big) + \frac{1}{2}\log\Big(\frac{N_2^2 + P_2}{N_2^2 + \theta_2^1 P_2}\Big). \tag{51}$$

However, because $\beta_2 > 0$ we have

$$R_2 = \frac{1}{2}\log\Big(\frac{N_2^2 + (\theta_2^1 + \theta_2^2)P_2}{N_2^2 + \theta_2^1 P_2}\Big). \tag{52}$$

Substituting from (52) into (51) and simplifying yields

$$R_0 = \frac{1}{2}\log\Big(\frac{N_1^2 + P_1}{N_1^2 + \theta_1^1 P_1}\Big) + \frac{1}{2}\log\Big(\frac{N_2^2 + P_2}{N_2^2 + (\theta_2^1 + \theta_2^2)P_2}\Big) > g_0(\theta), \tag{53}$$

which violates (17l) for any $\theta_2^2 > 0$. Hence, we must have $\beta_5 = 0$ which, using (50) and the fact that $N_1^2 > N_1^1$, leads to $\beta_6 = \beta_7 = 0$. We have thus shown that when $w_1 > w_2 > w_3$, $\beta_i = 0,\ i = 0, 4, \ldots, 11$, as desired.

## APPENDIX IV
## PROOF OF THEOREM 2

### *A. A solution of KKT system of equations* (17) *for original problem* (15) *for* $w_2 > w_1 > w_3$

For $w_2 > w_1 > w_3$, (17a)–(17c) yield

$$\beta_5 + \beta_{10} > \beta_1 \geq 0 \quad \text{and} \quad \beta_2 + \beta_5 + \beta_6 > \beta_9 \geq 0. \tag{54}$$

In Appendix III we showed that if $\beta_1 > 0$ then $\beta_5 = \beta_{10} = 0$, which contradicts the first inequality in (54). Hence, it is seen that in this case $\beta_1 = 0$. A similar argument can be used to show that $\beta_9 = 0$. We will show below that a solution of the KKT conditions in this case can be obtained by setting $\beta_2 = \beta_3 = 0$.

For $\beta_2 + \beta_3 = 0$ we have from (17e) that $\beta_0 = \beta_4 = \beta_8 = 0$. Applying this fact and the fact that $\beta_9 = 0$ to (17f) yields $\beta_{10} = \beta_{11} = 0$. Using

$$\beta_i = 0,\ i = 0, \ldots, 4, 8, \ldots, 11, \tag{55}$$

in (54), (17d)–(17g) and (17a)–(17c) with the current weight ordering yields $\beta_5 > 0$, $\beta_6 > 0$ and $\beta_7 > 0$, from which we have

$$R_0 + R_2 = g_{02}(\theta), \quad R_0 + R_1 + R_2 = g_{012}(\theta), \quad \text{and} \quad R_0 + R_1 + R_2 + R_3 = g_{0123}(\theta). \tag{56}$$

It remains to show that the rates generated in (56) are feasible, i.e., that setting $\beta_2 = \beta_3 = 0$ yields a solution of the KKT system of equations. Suppose that $\beta_2 + \beta_3 > 0$, then from (17e) we have that at least one $\{\beta_i\}_{i=0,4,8}$ is greater than zero. Using this fact in (17f) yields $\beta_{10} + \beta_{11} > 0$. Now, suppose that $\beta_5 + \beta_6 + \beta_7 = 0$. (This assumption results in fewer constraints being active.) In this case, we have $\beta_2 > 0, \beta_3 > 0, \beta_{10} > 0$ and $\beta_{11} > 0$. The complementarity slackness now implies that at least one of the constraints on $R_0$ is active, and for each rate $R_k$, $k = 1, 2, 3$, one can find two expressions. Equating these expressions, it is seen that in this case we have four equations in the four unknowns, $\{\theta_i^\ell\}_{i,\ell=1,2}$. Solving for these unknowns and substituting into (17d)–(17g) we have four equations in five unknowns, $\beta_2, \beta_3, \beta_{10}, \beta_{11}$ and one of $\{\beta_i\}_{i=0,4,8}$, in addition to the three equations (17a)–(17c). That is, in total we have seven linearly independent linear equations in five $\{\beta_i\}$ unknowns. Since these equations cannot be consistent we conclude that one must have $\beta_2 + \beta_3 = 0$ for the problem to be feasible.

### B. *A solution of KKT system of equations* (19) *for relaxed problem* (18) *for* $w_2 > w_1 > w_3$

For this weight ordering, the KKT conditions of the relaxed problem yield

$$\eta_{02} + \xi_{023} > \lambda_{01} \geq 0, \quad \text{and} \quad \lambda_{012} + \eta_{02} + \eta_{012} > \xi_{03} \geq 0. \tag{57}$$

Now, set

$$\lambda_{01} = \lambda_{012} = \lambda_{0123} = \xi_{03} = \xi_{023} = \xi_{0123} = 0. \tag{58}$$

Using this setting in (19a)–(19c), along with Lemma 4 and the fact that $w_2 > w_1 > w_3$, we have that

$$R_0 + R_2 = g_{02}(\alpha'), \quad R_0 + R_1 + R_2 = g_{012}(\alpha'), \quad R_0 + R_1 + R_2 + R_3 = g_{0123}(\alpha'). \tag{59}$$

Now, $\alpha$ and $\alpha''$ can be arbitrarily chosen so that (19h)–(19j) and (19p)–(19r) are satisfied, respectively.

### C. *Identifying solutions of* (17) *and* (19) *for* $w_2 > w_1 > w_3$

For $w_2 > w_1 > w_3$, we now compare the solution of the KKT system of the relaxed problem obtained in the previous section with that of the KKT system of the original problem in (15). Since a solution of (17) exists with $\beta_i = 0,\ i = 0, \ldots, 4, 8, \ldots, 11$, it is seen that for this solution (17d) and (17e) become identical to (27) and (28), respectively, when

$$\alpha' = \theta. \tag{60}$$

Now, we choose $\alpha$ and $\alpha''$ arbitrarily so that (19k)–(19r) are satisfied. Such $\alpha$ and $\alpha''$ exist because there exists a power partition $\theta$ that satisfies (17). Finally, it is seen that with $\alpha'$ identified with $\theta$, the rates generated by (42) are identical to those generated by (38), which completes the proof of Theorem 2.

# APPENDIX V
# PROOF OF THEOREM 3

We proceed by contradiction. In particular, we assume all the rates and partitions to be strictly positive and show that this leads to the desired contradiction, and that this contradiction is resolved by setting $R_2$ to zero.

## A. *Original problem with* $w_1 > w_3 > w_2$

In this section we show that, given $R_0$ satisfying (14) and a weight vector with $w_1 > w_3 > w_2$, any locally optimal solution of (15) must have $R_2 = 0$. First, assume that $R_k > 0$, for $k = 1, 2, 3$. We will show in this section that for this weight setting this assumption leads to a contradiction.

We begin by noting that under the assumption that $R_k > 0$, we have from (17a)–(17c) that

$$\beta_1 > \beta_5 + \beta_{10} \geq 0, \quad \text{and} \quad \beta_9 > \beta_2 + \beta_5 + \beta_6 \geq 0. \tag{61}$$

We will now show that $\beta_1$ and $\beta_9$ cannot be strictly greater than zero simultaneously, and we will use this to conclude that $R_2 = 0$. Since $\beta_1 > 0$, we have from (17e) that $\beta_2 + \beta_3 > 0$. Similarly, because $\beta_9 > 0$, we have from (17f) that $\beta_{10} + \beta_{11} > 0$. Also, since $\beta_1 > 0$, we have that

$$R_0 + R_1 = f_{01}(\theta), \quad \text{and since} \quad \beta_9 > 0, \tag{62}$$

$$R_0 + R_3 = h_{03}(\theta). \tag{63}$$

*1) Showing that* $\beta_3 = \beta_{11} = 0$*:* Suppose that $\beta_{11} > 0$. In this case, using (63) and the last equality in (17s), we have

$$R_1 + R_2 = \frac{1}{2}\log\Big(\frac{N_1^1 + \theta_1^1 P_1}{N_1^1}\Big) + \frac{1}{2}\log\Big(\frac{N_1^2 + (\theta_1^1 + \theta_1^2)P_1}{N_1^2 + \theta_1^1 P_1}\Big). \tag{64}$$

From (62) and (64), we have

$$\begin{aligned} R_0 &> R_0 + R_1 - (R_1 + R_2) \\ &= R_0 - R_2 \\ &= \frac{1}{2}\log\Big(\frac{N_1^1 + P_1}{N_1^1}\Big) + \frac{1}{2}\log\Big(\frac{N_2^3 + P_2}{N_2^3 + (\theta_2^1 + \theta_2^2)P_2}\Big) \end{aligned} \tag{65}$$

$$
\begin{aligned}
&- \frac{1}{2}\log\Big(\frac{N_1^1+\theta_1^1 P_1}{N_1^1}\Big) - \frac{1}{2}\log\Big(\frac{N_1^2+(\theta_1^1+\theta_1^2)P_1}{N_1^2+\theta_1^1 P_1}\Big)\\
&> \frac{1}{2}\log\Big(\frac{N_1^1+P_1}{N_1^1}\Big) + \frac{1}{2}\log\Big(\frac{N_2^3+P_2}{N_2^3+(\theta_2^1+\theta_2^2)P_2}\Big)\\
&- \frac{1}{2}\log\Big(\frac{N_1^1+\theta_1^1 P_1}{N_1^1}\Big) - \frac{1}{2}\log\Big(\frac{N_1^1+(\theta_1^1+\theta_1^2)P_1}{N_1^1+\theta_1^1 P_1}\Big)\\
&= \frac{1}{2}\log\Big(\frac{N_1^1+P_1}{N_1^1+(\theta_1^1+\theta_1^2)P_1}\Big) + \frac{1}{2}\log\Big(\frac{N_2^3+P_2}{N_2^3+(\theta_2^1+\theta_2^2)P_2}\Big),
\end{aligned}
\tag{66}
$$

which contradicts (17h). Hence we conclude that $\beta_{11}=0$. Notice that the contradiction here is resolved if $R_2=0$ and $\theta_1^2=0$. Using a similar argument, we conclude that $\beta_3=0$. The contradiction for the latter case is resolved if $R_2=0$ and $\theta_2^2=0$.

*2) Showing that $\beta_2=\beta_{10}=0$:* In order to show that $R_k$ cannot be strictly greater than zero for all $k=1,2,3$, we have to show that $\beta_2=0$ (or alternatively that $\beta_{10}=0$). This will contradict (17e) for $\beta_1>0$.

If $\beta_2>0$, then from the last equality in (17j) we have

$$
R_0+R_1+R_2=f_{012}(\theta).
$$

Using this in the middle inequality in (17k), we have $R_3 \le \frac{1}{2}\log\Big(\frac{N_2^1+\alpha_2^1 P_2}{N_2^1}\Big)$. Using that in (63) yields,

$$
\begin{aligned}
R_0 &\ge \frac{1}{2}\log\Big(\frac{N_1^3+P_1}{N_1^3+(\theta_1^1+\theta_1^2)P_1}\Big) + \frac{1}{2}\log\Big(\frac{N_2^1+P_2}{N_2^1}\Big) - \frac{1}{2}\log\Big(\frac{N_2^1+\theta_2^1 P_2}{N_2^1}\Big)\\
&= \frac{1}{2}\log\Big(\frac{N_1^3+P_1}{N_1^3+(\theta_1^1+\theta_1^2)P_1}\Big) + \frac{1}{2}\log\Big(\frac{N_2^1+P_2}{N_2^1+\theta_2^1 P_2}\Big).
\end{aligned}
$$

This inequality contradicts (17p) for any $\theta_2^2>0$, and the contradiction is resolved if $\theta_2^2=0$. The argument that $\beta_{10}=0$ can be made analogously, and the contradiction in that case is resolved if $\theta_1^2=0$.

### B. Relaxed problem with $w_1>w_3>w_2$

In this section we show that for $w_1>w_3>w_2$, the optimal solution of (18) must have $R_2=0$. To show this, we will assume that $R_k>0$, for $k=1,2,3$, and then proceed by contradiction. If $R_k>0$, for $k=1,2,3$, then from (19a) and (19c) we have that

$$
\lambda_{01}+\lambda_{012}+\eta_{012} > \xi_{03}+\xi_{023} \ge 0, \tag{67}
$$

and from (19b) and (19c), we have that

$$
\xi_{03} > \eta_{02}+\eta_{012}+\lambda_{012} \ge 0, \tag{68}
$$

from which we have

$$\lambda_{01} > \eta_{02} + \xi_{023} \geq 0. \tag{69}$$

Now, using (68) and (69), we have from (29), (31), (35), (19h)–(19j) and (19p)–(19r) that

$$R_0 + R_1 = f_{01}(\alpha), \quad R_2 = \frac{1}{2}\log\Big(\frac{N_2^2 + (\alpha_2^1 + \alpha_2^2)P_2}{N_2^2 + \alpha_2^1 P_2}\Big) \quad \text{and} \quad R_3 = \frac{1}{2}\log\Big(\frac{N_2^1 + \alpha_2^1 P_2}{N_2^1}\Big), \tag{70}$$

and

$$R_0 + R_3 = h_{03}(\alpha''), \quad R_2 = \frac{1}{2}\log\Big(\frac{N_1^2 + (\alpha_1''^1 + \alpha_1''^2)P_1}{N_1^2 + \alpha_1''^1 P_1}\Big) \quad \text{and} \quad R_1 = \frac{1}{2}\log\Big(\frac{N_1^1 + \alpha_1''^1 P_1}{N_1^1}\Big). \tag{71}$$

We will now show that $R_2 = 0$ and that $\alpha_1''^2 = \alpha_2^2 = 0$. Consider the value of the objective that corresponds to the equalities in (70), namely

$$w_1\Big(\frac{1}{2}\log\Big(\frac{N_1^1 + P_1}{N_1^1}\Big) + \frac{1}{2}\log\Big(\frac{N_2^3 + P_2}{N_2^3 + (\alpha_2^1 + \alpha_2^2)P_2}\Big) - R_0\Big) + w_2\frac{1}{2}\log\Big(\frac{N_2^2 + (\alpha_2^1 + \alpha_2^2)P_2}{N_2^2 + \alpha_2^1 P_2}\Big) + w_3\frac{1}{2}\log\Big(\frac{N_2^1 + \alpha_2^1 P_2}{N_2^1}\Big). \tag{72}$$

Observe that, for a given $R_0$, the value of the objective does not depend on the partitions $\alpha_1^1$ and $\alpha_1^2$ Consider now the value of the objective that corresponds to power partitions $\gamma_2^1 = \alpha_2^1 + \alpha_2^2$ and $\gamma_2^2 = 0$. In this case the value of the objective is given by

$$w_1\Big(\frac{1}{2}\log\Big(\frac{N_1^1 + P_1}{N_1^1}\Big) + \frac{1}{2}\log\Big(\frac{N_2^3 + P_2}{N_2^3 + (\alpha_2^1 + \alpha_2^2)P_2}\Big) - R_0\Big) + w_3\frac{1}{2}\log\Big(\frac{N_2^1 + (\alpha_2^1 + \alpha_2^2)P_2}{N_2^1}\Big). \tag{73}$$

Subtracting (72) from (73), we have

$$w_3\frac{1}{2}\log\Big(\frac{N_2^1 + (\alpha_2^1 + \alpha_2^2)P_2}{N_2^1 + \alpha_2^1 P_2}\Big) - w_2\frac{1}{2}\log\Big(\frac{N_2^2 + (\alpha_2^1 + \alpha_2^2)P_2}{N_2^2 + \alpha_2^1 P_2}\Big) > 0$$

where the inequality follows from the fact that $N_2^2 > N_2^1$ and $w_2 < w_3$. Hence, it is seen that, for this weight ordering, the rates yielded by (70) are not optimal unless $R_2 = 0$ and $\alpha_2^2 = 0$. Using a similar argument, we can show that the rates yielded by (71) are not optimal unless $R_2 = 0$ and $\alpha_1''^2 = 0$. Observe that the rates yielded by setting $R_2 = 0$ and $\alpha_1''^2 = \alpha_2^2 = 0$ are feasible because these rates are feasible for the restricted case of $\alpha = \alpha''$; see Appendix V-A.

## Appendix VI

## The case of $w_2 > w_3 > w_1$

In this section we show that in the case of $w_2 > w_3 > w_1$ the rates and partitions generated by solving the KKT system in (17) corresponding to (15) does not belong to the feasible set of (16), and hence does not belong to the intersection of the two regions, which is the SPC region of interest. For this ordering, because of the symmetry between the degradation order of receivers $Y$ and $W$ (cf. Figure 1), feasible rates and partitions can be generated by solving the KKT system corresponding to the description of Region$_2$ in (16) in way similar to the one used to generate feasible rates and partitions for $w_2 > w_1 > w_3$ from the KKT system corresponding to the description of Region$_1$ in (15).

We now analyze the KKT conditions for this particular ordering. Using $w_2 > w_3 > w_1$ in (19a)–(19c), we have

$$\beta_2 + \beta_5 + \beta_6 > \beta_9 \geq 0 \qquad \text{and} \qquad \beta_9 + \beta_{10} > \beta_1 \geq 0, \tag{74}$$

from which we have

$$\beta_5 + \beta_{10} > \beta_1 \geq 0. \tag{75}$$

As argued in Appendix IV, if $\beta_1 > 0$ then $\beta_5 = \beta_{10} = 0$, which contradicts the second inequality in (75). Similarly, if $\beta_9 > 0$ then $\beta_2 = \beta_5 = \beta_6 = 0$, which contradicts the first inequality in (74). Hence, we conclude that $\beta_1 = \beta_9 = 0$. Since $\beta_9 = 0$, we have from the second inequality in (74) that

$$\beta_{10} > 0. \tag{76}$$

Using the first inequality in (74) in (17d) yields

$$\beta_3 + \beta_7 > 0. \tag{77}$$

Using (76) in (17g) yields

$$\beta_6 + \beta_7 + \beta_{11} > 0. \tag{78}$$

We now consider all possible assumptions for $\beta_5, \beta_6$ and $\beta_7$:

- $[\beta_5 + \beta_6 + \beta_7 = 0]$, $[\beta_5 > 0,\ \beta_6 > 0, \beta_7 > 0]$, $[\beta_7 = 0,\ \beta_5 > 0, \beta_6 > 0]$ and $[\beta_6 + \beta_7 = 0,\ \beta_5 > 0]$: These assumptions can be eliminated by using an argument similar to the one in Appendix IV-A to show that they result in a number of independent linear equations that exceeds the number of unknowns;
- $[\beta_5 + \beta_6 = 0,\ \beta_7 > 0]$: For this assumption, the first inequality in (74) yields $\beta_2 > 0$. If $R_3 < \frac{1}{2}\log\left(\frac{N_2^1 + \theta_2^1 P_2}{N_2^1}\right)$, which is possible because we have no conditions on $\beta_3$, the fact that $\beta_7 > 0$ implies that $R_0 + R_1 + R_2 > g_{012}(\theta)$, which violates (17n). Now, if $R_3 = \frac{1}{2}\log\left(\frac{N_2^1 + \theta_2^1 P_2}{N_2^1}\right)$, we will have a number of independent equations that exceeds the number of unknowns. Hence, this assumption can be eliminated;
- $[\beta_5 + \beta_7 = 0,\ \beta_6 > 0]$: This assumption yields $\beta_3 > 0$. If $R_3 < \frac{1}{2}\log\left(\frac{N_2^1 + \theta_2^1 P_2}{N_2^1}\right)$, which is possible because under this assumption $\beta_7 = 0$, the fact that $\beta_3 > 0$ implies that $R_0 + R_1 + R_2 > f_{012}(\theta)$, which violates (17j). Now, if $R_3 = \frac{1}{2}\log\left(\frac{N_2^1+\theta_2^1 P_2}{N_2^1}\right)$, we will have a number of independent linear equations that exceeds the number of unknowns. Hence, this assumption can be eliminated;

- $[\beta_5 = 0,\ \beta_6 > 0, \beta_7 > 0]$: For this assumption, $R_3 = \frac{1}{2}\log\left(\frac{N_2^1+\theta_2^1 P_2}{N_2^1}\right)$ and $R_0 + R_1 + R_2 = g_{012}(\theta)$. If $R_1 < \frac{1}{2}\log\left(\frac{N_1^1+\theta_1^1 P_1}{N_1^1}\right)$, we have $R_0 + R_2 > g_{02}(\theta)$, which violates (17m), and if $R_1 = \frac{1}{2}\log\left(\frac{N_1^1+\theta_1^1 P_1}{N_1^1}\right)$, the number of independent linear equations exceeds the number of unknowns. Hence, this assumption can be eliminated;
- $[\beta_6 = 0,\ \beta_5 > 0, \beta_7 > 0]$: For this assumption we have

$$R_0 + R_2 = g_{02}(\theta), \qquad \text{and} \tag{79}$$

$$R_0 + R_1 + R_2 + R_3 = g_{0123}(\theta). \tag{80}$$

Now, if $R_1 = \frac{1}{2}\log\left(\frac{N_1^1+\theta_1^1 P_1}{N_1^1}\right)$, one can see that the number of independent linear equations exceeds the number of unknowns. Hence, we have $R_1 < \frac{1}{2}\log\left(\frac{N_1^1+\theta_1^1 P_1}{N_1^1}\right)$. Using this in (80) yields $R_3 > \frac{1}{2}\log\left(\frac{N_2^1+\theta_2^1 P_2}{N_2^1}\right)$. That, together with (79), implies that $R_0 + R_2 + R_3 > g_{023}(\theta)$. This implies, in turn, that this solution, although feasible for (15), is not feasible for (16) and hence is not in the SPC region, which establishes the desired result.

## Appendix VII
## A convex transformation of (18)

In this section we will transform the relaxed problem in (18) into a convex form. In particular, we will show that this problem can be a cast as a geometric program. We will assume that the powers $P_1$ and $P_2$ are given. However, the methodology that we use can be extended to the case in which the powers are not fixed *a priori*. In order to perform the required transformation, we use the following change of variables

$$t_k = e^{2R_k},\ k = 0, \ldots, 3 \tag{81}$$
$$Q_i^j = \alpha_i^j P_i, \quad Q_i'^j = \alpha_i'^j P_i, \quad Q_i''^j = \alpha_i''^j P_i.$$

Now, using the monotonicity of the log function, the optimization problem in (18) can be cast as

$$\max \qquad t_1^{w_1} t_2^{w_2} t_3^{w_3} \tag{82a}$$

subject to

$$t_0(N_1^1 + Q_1^1 + Q_1^2)(N_2^3 + Q_2^1 + Q_2^2)(P_1 + N_1^1)^{-1}(P_2 + N_2^3)^{-1} \le 1, \tag{82b}$$

$$N_1^1 t_0 t_1 (N_2^3 + Q_2^1 + Q_2^2)(P_1 + N_1^1)^{-1}(P_2 + N_2^3)^{-1} \le 1, \tag{82c}$$

$$N_1^1 t_0 t_1 t_2 (N_2^3 + Q_2^1 + Q_2^2)(N_2^2 + Q_2^1)(P_1 + N_1^1)^{-1}(P_2 + N_2^3)^{-1}(N_2^2 + Q_2^1 + Q_2^2)^{-1} \leq 1, \tag{82d}$$

$$N_2^1 N_1^1 t_0 t_1 t_2 t_3 (N_2^3 + Q_2^1 + Q_2^2)(N_2^2 + Q_2^1)(P_1 + N_1^1)^{-1}(P_2 + N_2^3)^{-1}(N_2^1 + Q_2^1)^{-1}(N_2^2 + Q_2^1 + Q_2^2)^{-1} \leq 1, \tag{82e}$$

$$t_0 (N_1^2 + Q_1'^1 + Q_1'^2)(P_1 + N_1^2)^{-1}(N_2^2 + Q_2'^1 + Q_2^{2\prime})(P_2 + N_2^2)^{-1} \leq 1, \tag{82f}$$

$$t_0 t_2 (N_1^2 + Q_1'^1)(N_2^2 + Q_2'^1)(P_1 + \Delta_1^2)^{-1}(P_2 + N_2^2)^{-1} \leq 1, \tag{82g}$$

$$N_1^1 t_0 t_1 t_2 (N_1^2 + Q_1'^1)(P_1 + \Delta_1^2)^{-1}(P_2 + \Delta_2^2)^{-1}(N_2^2 + Q_2'^1)(N_1^1 + Q_1'^1)^{-1} \leq 1, \tag{82h}$$

$$N_2^1 N_1^1 t_0 t_1 t_2 t_3 (N_1^2 + Q_1'^1)(N_2^2 + Q_2'^1)(N_1^1 + Q_1'^1)^{-1}(N_2^1 + Q_2'^1)^{-1}(P_2 + N_2^2)^{-1}(P_1 + N_1^2)^{-1} \leq 1, \tag{82i}$$

$$t_0 (N_1^3 + Q_1''^1 + Q_1''^2)(P_1 + N_1^3)^{-1}(N_2^1 + Q_2''^1 + Q_2''^2)(P_2 + N_2^1)^{-1} \leq 1, \tag{82j}$$

$$N_2^1 t_0 t_3 (N_1^3 + Q_1''^1 + Q_1''^2)(P_1 + N_1^3)^{-1}(P_2 + N_2^1)^{-1} \leq 1, \tag{82k}$$

$$N_2^1 t_0 t_2 t_3 (N_1^3 + Q_1''^2 + Q_1''^1)(N_1^2 + Q_1''^1)(P_1 + N_1^3)^{-1}(P_2 + N_2^1)^{-1}(N_1^2 + Q_1''^1 + Q_1''^2)^{-1} \leq 1, \tag{82l}$$

$$N_2^1 N_1^1 t_0 t_1 t_2 t_3 (N_1^3 + Q_1''^1 + Q_1''^2)(N_1^2 + Q_1''^1)(P_1 + N_1^3)^{-1}(P_2 + N_2^1)^{-1}(N_1^1 + Q_1''^1)^{-1} \times (N_1^2 + Q_1''^1 + Q_1''^2)^{-1} \leq 1, \tag{82m}$$

$$\sum_{\ell=1}^{2} Q_i^\ell \leq P_i, \qquad i = 1, 2, \tag{82n}$$

$$\sum_{\ell=1}^{2} Q_i'^\ell \leq P_i, \qquad i = 1, 2, \tag{82o}$$

$$\sum_{\ell=1}^{2} Q_i''^\ell \leq P_i, \qquad i = 1, 2, \tag{82p}$$

$$Q_i^\ell \geq 0, \quad Q_i'^\ell \geq 0, \quad Q_i''^\ell \geq 0, \quad i = 1, 2, \ \ell = 1, 2, 3, \tag{82q}$$

Let

$$\begin{aligned}
T_1^1 &= Q_1^1 + N_1^1/2, & T_1^2 &= Q_1^2 + N_1^1/2, \\
T_2^1 &= Q_2^1 + N_2^1, & T_2^2 &= Q_2^1 + Q_2^2 + N_2^2, \\
T_1'^1 &= Q_1'^1 + N_1^2, & T_1'^2 &= Q_1'^2 + N_1^2 - N_1^1, \\
T_2'^1 &= Q_2'^1 + N_2^2, & T_2'^2 &= Q_2'^2 + N_2^2 - N_2^1, \\
T_1''^1 &= Q_1''^1 + N_1^1, & T_1''^2 &= Q_1''^1 + Q_1''^2 + N_1^2, \\
T_2''^1 &= Q_2''^1 + N_2^1/2, \quad \text{and} & T_2''^2 &= Q_2''^2 + N_2^1/2.
\end{aligned}$$

Hence,

$$Q_1^1 = T_1^1 - N_1^1/2, \qquad Q_1^2 = T_1^2 - N_1^1/2, \tag{83a}$$

$$Q_2^1 = T_2^1 - N_2^1, \qquad Q_2^2 = T_2^2 - T_2^1 + N_2^1 - N_2^2, \tag{83b}$$

$$Q_1'^1 = T_1'^1 - N_1^2, \qquad Q_1'^2 = T_1'^2 - (N_1^2 - N_1^1), \tag{83c}$$

$$Q_2'^1 = T_2'^1 - N_2^2, \qquad Q_2'^2 = T_2'^2 - (N_2^2 - N_2^1), \tag{83d}$$

$$Q_1''^1 = T_1''^1 - N_1^1, \qquad Q_1''^2 = T_1''^2 - T_1''^1 + N_1^1 - N_1^2, \tag{83e}$$

$$Q_2''^1 = T_2''^1 - N_2^1/2, \quad \text{and} \quad Q_2''^2 = T_2''^2 - N_2^1/2. \tag{83f}$$

Using these new variables, we can re-write (82n) as

$$\sum_{\ell=1}^{2} T_1^\ell \leq P_1 + N_1^1, \quad T_2^2 \leq P_2 + N_2^2. \tag{84}$$

The constraints in (82o) can be re-written as

$$T_i'^1 + T_i'^2 \leq P_i + 2N_i^2 - N_i^1, \quad i = 1, 2, \tag{85}$$

and the constraints in (82p) can be re-written as

$$T_1''^2 \leq P_1 + N_1^2, \quad \sum_{\ell=1}^{2} T_2''^\ell \leq P_2 + N_2^1. \tag{86}$$

We now consider the conditions in (82q). First we note that by replacing the equalities in (82n)–(82p) by the inequalities in (84)–(86), $Q_i^3$, $Q_i'^3$, and $Q_i''^3$ are eliminated from the formulation. (These variables do not appear in any other constraint.) For the first set of constraints in (82q), we have

$$T_1^1 \geq N_1^1/2, \quad T_1^2 \geq N_1^1/2, \quad T_2^1 \geq N_2^1, \quad \text{and} \quad T_2^2 \geq T_2^1 + N_2^2 - N_2^1. \tag{87}$$

For the second set, we have

$$T_i'^1 \geq N_i^2, \quad \text{and} \quad T_i'^2 \geq N_i^2 - N_i^1, \quad i = 1, 2. \tag{88}$$

For the last set, we have

$$T_1''^1 \geq N_1^1, \quad T_1''^2 \geq T_1''^1 + N_1^2 - N_1^1, \quad \text{and} \quad T_2''^\ell \geq N_2^1/2. \tag{89}$$

Before proceeding to show how the remaining constraints can be cast as a geometric program, we recall that the degradedness condition $N_i^{\ell+1} > N_i^\ell$, for $i, \ell = 1, 2$. Hence, one can see that all the transformed constraints in (84)–(89) are in the form of posynomials that can be readily incorporated in a (convex) geometric program. Using the transformation in (83), we can write the constraints in (82b)–(82e) as follows:

$$t_0(T_1^1 + T_1^2)(N_2^3 - N_2^2 + T_2^2)(P_1 + N_1^1)^{-1}(P_2 + N_2^3)^{-1} \leq 1, \tag{90a}$$

$$N_1^1 t_0 t_1 (N_2^3 - N_2^2 + T_2^2)(P_1 + N_1^1)^{-1}(P_2 + N_2^3)^{-1} \leq 1, \tag{90b}$$

$$N_1^1 t_0 t_1 t_2 (N_2^3 - N_2^2 + T_2^2)(N_2^2 - N_2^1 + T_2^1)(P_1 + N_1^1)^{-1}(P_2 + N_2^3)^{-1} \leq T_2^2, \tag{90c}$$

$$N_2^1 N_1^1 t_0 t_1 t_2 t_3 (N_2^3 - N_2^2 + T_2^2)(N_2^2 - N_2^1 + T_2^1)(P_1 + N_1^1)^{-1}(P_2 + N_2^3)^{-1} \leq T_2^1 T_2^2. \tag{90d}$$

Note that because $N_2^2 - N_2^1 > 0$, all the constraints in (90a)–(90d) are in the standard posynomial form. Consider now the constraints in (82f)–(82i). Using the transformations in (83), these constraints can be written as

$$t_0(T_1'^1 + T_1'^2)(P_1 + N_1^2)^{-1}(T_2'^1 + T_2^{2\prime})(P_2 + N_2^2)^{-1} \leq 1, \tag{91a}$$

$$t_0 t_2 (N_1^2 - N_1^1 + T_1'^1)(N_2^2 - N_2^1 + T_2'^1)(P_1 + N_1^2)^{-1}(P_2 + N_2^2)^{-1} \leq 1, \tag{91b}$$

$$N_1^1 t_0 t_1 t_2 (N_1^2 - N_1^1 + T_1'^1)(P_1 + N_1^2)^{-1}(P_2 + N_2^2)^{-1}(N_2^2 - N_2^1 + T_2'^1) \leq T_1'^1, \tag{91c}$$

$$N_2^1 N_1^1 t_0 t_1 t_2 t_3 (N_1^2 - N_1^1 + T_1'^1)(N_2^2 - N_2^1 + T_2'^1)(P_1 + N_1^2)^{-1}(P_2 + N_2^2)^{-1} \leq T_1'^1 T_2'^1. \tag{91d}$$

One can also see that (91a)–(91d) are in the form of posynomial constraints. Finally, we express the constraints in (82j)–(82m) as

$$t_0(N_1^3 - N_1^2 + T_1''^2)(T_2''^1 + T_2''^2)(P_1 + N_1^3)^{-1}(P_2 + N_2^1)^{-1} \leq 1, \tag{92a}$$

$$N_2^1 t_0 t_3 (N_1^3 - N_1^2 + T_1''^2)(P_1 + N_1^3)^{-1}(P_2 + N_2^1)^{-1} \leq 1, \tag{92b}$$

$$N_2^1 t_0 t_2 t_3 (N_1^3 - N_1^2 + T_1''^2)(N_1^2 - N_1^1 + T_1''^1)(P_1 + N_1^3)^{-1}(P_2 + N_2^1)^{-1} \leq T_1''^2, \tag{92c}$$

$$N_2^1 N_1^1 t_0 t_1 t_2 t_3 (N_1^3 - N_1^2 + T_1''^2)(N_1^2 - N_1^1 + T_1''^1)(P_1 + N_1^3)^{-1}(P_2 + N_2^1)^{-1} \leq T_1''^1 T_1''^2. \tag{92d}$$

Seeing as (92a)–(92d) are in the form of posynomial constraints, we can now write (82) as

$$\max \quad t_1^{w_1} t_2^{w_2} t_3^{w_3} \tag{93a}$$

$$\text{subject to} \quad \text{(90a)–(90d)} \tag{93b}$$

$$\text{(91a)–(91d)} \tag{93c}$$

$$\text{(92a)–(92d)} \tag{93d}$$

$$\text{(84)–(89).} \tag{93e}$$

Since the objective is in the form of a monomial and all the constraints are in the form of posynomials, the problem in (93) is readily seen to be a geometric program.

We now transform this geometric program into a convex form. In order to do that, we take the logarithm of the objective and the constraints in (93), and we use the transformations

$$X_i^\ell = \log(T_i^\ell), \quad X_i'^\ell = \log(T_i'^\ell), \quad \text{and} \quad X_i''^\ell = \log(T_i''^\ell), \qquad i = 1,2, \quad \ell = 1,2. \tag{94}$$

We will also use (81) to write $\log(t_k) = 2R_k,\ k = 0,\dots,3$. Using these transformations, the problem in (93) can be written as

$$\max \sum_{k=1}^{3} w_k R_k \tag{95a}$$

$$\text{subject to } 2R_0 \le \log\Big(\frac{N_1^1 + P_1}{e^{X_1^1} + e^{X_1^2}}\Big) + \log\Big(\frac{N_2^3 + P_2}{N_2^3 - N_2^2 + e^{X_2^2}}\Big) \tag{95b}$$

$$2R_0 + 2R_1 \le \log\Big(\frac{N_1^1 + P_1}{N_1^1}\Big) + \log\Big(\frac{N_2^3 + P_2}{N_2^3 - N_2^2 + e^{X_2^2}}\Big) \tag{95c}$$

$$2R_0 + 2R_1 + 2R_2 \le \log\Big(\frac{N_1^1 + P_1}{N_1^1}\Big) + \log\Big(\frac{N_2^3 + P_2}{N_2^3 - N_2^2 + e^{X_2^2}}\Big) + \log\Big(\frac{e^{X_2^2}}{N_2^2 - N_2^1 + e^{X_2^1}}\Big) \tag{95d}$$

$$2R_0 + 2R_1 + 2R_2 + 2R_3 \le \log\Big(\frac{N_1^1 + P_1}{N_1^1}\Big) + \log\Big(\frac{N_2^3 + P_2}{N_2^3 - N_2^2 + e^{X_2^2}}\Big) + \log\Big(\frac{e^{X_2^2}}{N_2^2 - N_2^1 + e^{X_2^1}}\Big) + \log\Big(\frac{e^{X_2^1}}{N_2^1}\Big) \tag{95e}$$

$$2R_0 \le \log\Big(\frac{N_1^2 + P_1}{e^{X_1'^1} + e^{X_1'^2}}\Big) + \log\Big(\frac{N_2^2 + P_2}{e^{X_2'^1} + e^{X_2'^2}}\Big) \tag{95f}$$

$$2R_0 + 2R_2 \le \log\Big(\frac{N_1^2 + P_1}{N_1^2 - N_1^1 + e^{X_1'^1}}\Big) + \log\Big(\frac{N_2^2 + P_2}{N_2^2 - N_2^1 + e^{X_2'^1}}\Big) \tag{95g}$$

$$2R_0 + 2R_1 + 2R_2 \le \log\Big(\frac{N_1^2 + P_1}{N_1^2 - N_1^1 + e^{X_1'^1}}\Big) + \log\Big(\frac{N_2^2 + P_2}{N_2^2 - N_2^1 + e^{X_2'^1}}\Big) + \log\Big(\frac{e^{X_1'^1}}{N_1^1}\Big) \tag{95h}$$

$$2R_0 + 2R_1 + 2R_2 + 2R_3 \le \log\Big(\frac{N_1^2 + P_1}{N_1^2 - N_1^1 + e^{X_1'^1}}\Big) + \log\Big(\frac{N_2^2 + P_2}{N_2^2 - N_2^1 + e^{X_2'^1}}\Big) + \log\Big(\frac{e^{X_1'^1}}{N_1^1}\Big) + \log\Big(\frac{e^{X_2'^1}}{N_2^1}\Big) \tag{95i}$$

$$2R_0 \le \log\Big(\frac{N_1^3 + P_1}{N_1^3 - N_1^2 + e^{X_1''^1}}\Big) + \log\Big(\frac{N_2^1 + P_2}{e^{X_2''^1} + e^{X_2''^2}}\Big) \tag{95j}$$

$$2R_0 + 2R_3 \le \log\Big(\frac{N_1^3 + P_1}{N_1^3 - N_1^2 + e^{X_1''^1}}\Big) + \log\Big(\frac{N_2^1 + P_2}{N_2^1}\Big) \tag{95k}$$

$$2R_0 + 2R_2 + 2R_3 \leq \log\Big(\frac{N_1^3 + P_1}{N_1^3 - N_1^2 + e^{X_1''^1}}\Big) + \log\Big(\frac{N_2^1 + P_2}{N_2^1}\Big) + \log\Big(\frac{e^{X_1''^2}}{N_1^2 - N_1^1 + e^{X_1''^1}}\Big) \tag{95l}$$

$$2R_0 + 2R_1 + 2R_2 + 2R_3 \leq \log\Big(\frac{N_1^3 + P_1}{N_1^3 - N_1^2 + e^{X_1''^1}}\Big) + \log\Big(\frac{N_2^1 + P_2}{N_2^1}\Big) + \log\Big(\frac{e^{X_1''^2}}{N_1^2 - N_1^1 + e^{X_1''^1}}\Big) + \log\Big(\frac{e^{X_1'^1}}{N_1^1}\Big) \tag{95m}$$

$$\sum_{\ell=1}^{2} e^{X_1^\ell} \leq P_1 + N_1^1, \qquad e^{X_2^2} \leq P_2 + N_2^2, \tag{95n}$$

$$e^{X_i'^1} + e^{X_i'^2} \leq P_i + 2N_i^2 - N_i^1, \quad i = 1, 2, \tag{95o}$$

$$e^{X_1''^2} \leq P_1 + N_1^2, \qquad \sum_{\ell=1}^{2} e^{X_2''^\ell} \leq P_2 + N_2^1, \tag{95p}$$

$$e^{X_1^1} \geq N_1^1/2, \quad e^{X_1^2} \geq N_1^1/2, \quad e^{X_2^1} \geq N_2^1, \qquad e^{X_2^2} \geq e^{X_2^1} + N_2^2 - N_2^1, \tag{95q}$$

$$e^{X_i'^1} \geq N_i^2, \quad e^{X_i'^2} \geq N_i^2 - N_i^1, \qquad i = 1, 2, \tag{95r}$$

$$e^{X_1''^1} \geq N_1^1, \quad e^{X_1''^2} \geq e^{X_1''^1} + N_1^2 - N_1^1, \quad e^{X_2''^\ell} \geq N_2^1/2, \quad \ell = 1, 2. \tag{95s}$$

Note that this problem is identical to (18), but with the power partitions parameterized by the exponential function.

## APPENDIX VIII
## PROOF OF THEOREM 4

First, we note that for any $P_1$ and $P_2$ greater than zero, the problem in (18) is strictly feasible. From Section VI it is seen that for each weight ordering, the active constraints at the provided solutions are linearly independent. (For each weight ordering, each constraint that is active at the provided solution involves a distinct partial sum of $\{R_k\}_{k=1}^3$.) Hence, using Proposition 3.3.1 in [27], it is seen that the KKT conditions are necessary for optimality. We now show that these conditions are also sufficient. In order to do that, we use [27, Proposition 5.1.5]. Let $L(\mathcal{A}, \gamma)$ denote the Lagrangian function at the vector of primal variables, $\mathcal{A}$, and the Lagrange multipliers, $\gamma$. Then, from [27, Proposition 5.1.5] it is seen that it is sufficient to show that, for any vector $\gamma \geq 0$, if the vector $\mathcal{A}^* \triangleq (R_1^*, R_2^*, R_3^*, \alpha^*, \alpha'^*, \alpha''^*)$ satisfies $\nabla_{\mathcal{A}} L(\mathcal{A}, \gamma)|_{\mathcal{A}=\mathcal{A}^*} = 0$, then it maximizes $L(\mathcal{A}, \gamma)$ for all feasible vectors $\mathcal{A}$. In order to show this, we recall that in Appendix VII we showed that (18) can be transformed into the convex form in (95). Let $L_c$ be the Lagrangian function that corresponds to this convex problem, and let $\mathcal{B}$ be the vector of transformed variables in (94). Now,

$$\nabla_{\mathcal{A}} L(\mathcal{A},\gamma) = J \nabla_{\mathcal{B}} L_c(\mathcal{B},\gamma), \tag{96}$$

where $J$ is the Jacobian matrix of the transformation in (94), i.e., the $ij$-th entry of $J$ is given by $\frac{\partial \mathcal{B}_i}{\partial \mathcal{A}_j}$. First we notice that this transformation is continuous, one-to-one and invertible. Now, one can easily check that

$$J = \begin{bmatrix} I_3 & 0 & 0 & 0 \\ 0 & J_1 & 0 & 0 \\ 0 & 0 & J_2 & 0 \\ 0 & 0 & 0 & J_3 \end{bmatrix}, \tag{97}$$

where $I_3$ is the $3 \times 3$ identity matrix, and

$$J_1 = \begin{bmatrix} \frac{P_1}{N_1^1/2+P_1\alpha_1^1} & 0 & 0 & 0 \\ 0 & \frac{P_1}{N_1^1/2+P_1\alpha_1^2} & 0 & 0 \\ 0 & 0 & \frac{P_2}{N_2^1+P_2\alpha_2^1} & 0 \\ 0 & 0 & \frac{P_2}{N_2^1+P_2\alpha_2^1+P_2\alpha_2^2} & \frac{P_2}{N_2^1+P_2\alpha_2^1+P_2\alpha_2^2} \end{bmatrix},$$

$$J_2 = \begin{bmatrix} \frac{P_1}{N_1^2+P_1\alpha_1'^1} & 0 & 0 & 0 \\ 0 & \frac{P_1}{N_1^2-N_1^1+P_1\alpha_1'^2} & 0 & 0 \\ 0 & 0 & \frac{P_2}{N_2^2+P_2\alpha_2'^1} & 0 \\ 0 & 0 & 0 & \frac{P_2}{N_2^2-N_2^1+P_2\alpha_2'^2} \end{bmatrix},$$

$$J_3 = \begin{bmatrix} \frac{P_1}{N_1^1+P_1\alpha_1''^1} & 0 & 0 & 0 \\ \frac{P_1}{N_1^2+P_1\alpha_1''^1+P_1\alpha_1''^2} & \frac{P_1}{N_1^2+P_1\alpha_1''^1+P_1\alpha_1''^2} & 0 & 0 \\ 0 & 0 & \frac{P_2}{N_2^1/2+P_2\alpha_2''^1} & 0 \\ 0 & 0 & 0 & \frac{P_2}{N_2^1/2+P_2\alpha_2''^2} \end{bmatrix}.$$

It is clear from (97) that for any $\mathcal{A} \geq 0$, the matrix $J$ is non-singular. Together with (96) this implies that $\nabla_{\mathcal{A}} L(\mathcal{A},\gamma) = 0$ if and only if $\nabla L_c(\mathcal{B},\gamma) = 0$. The convexity of the problem in (82) implies that $\nabla_{\mathcal{B}} L_c(\mathcal{B},\gamma) = 0$ only at the global maximum of $L_c(\mathcal{B},\gamma)$. Hence, from the continuity and the one-to-one correspondence of the transformation in (94), one can see that $\nabla_{\mathcal{A}} L(\mathcal{A},\gamma)$ equals zero only at the global maximum of $L(\mathcal{A},\gamma)$, for any given vector $\gamma \geq 0$, and hence for the optimal Lagrange multipliers $\gamma^*$.

# APPENDIX IX
## PROOF OF THE INEQUALITIES IN (22)

Given some small positive reals $\epsilon_j,\ j=1,2,3$, for every achievable rate, there is a sufficiently large $n$ such that $P_{e_j}^n < \epsilon_j$. It follows from Fano's inequality that

$$H(M_0, M_1|Y_1, Y_2) \le n\epsilon_1, \tag{98a}$$

$$H(M_0, M_2|Z_1, Z_2) \le n\epsilon_2, \tag{98b}$$

$$H(M_0, M_3|W_1, W_2) \le n\epsilon_3. \tag{98c}$$

In order obtain (22a), we have from Fano's inequality that

$$\begin{aligned}
nR_0 = H(M_0) &\le I(M_0; Y_1, Y_2) + n\epsilon_1 \\
&\le I(M_0; Y_2) + I(M_0; Y_1|Y_2) + n\epsilon_1 \\
&\le I(M_0; Y_2) + I(M_0, Y_2, Z_2; Y_1) + n\epsilon_1 \\
&\le I(M_0, M_1, Z_1, W_1; Y_2) + I(M_0, M_3, Z_2, Y_2; Y_1) + n\epsilon_1 \\
&= I(\mathcal{U}_1^3; Y_1) + I(\mathcal{U}_2^3; Y_2) + n\epsilon_1,
\end{aligned} \tag{99}$$

where $\mathcal{U}_1^3$ and $\mathcal{U}_2^3$ are defined in (21). Due to the symmetry between receivers $Y$ and $W$, (22e) can be proved in a similar manner.

We now show how to obtain the bound in (22b). Using Fano's inequality, we have

$$\begin{aligned}
n(R_0 + R_1) &\le I(M_0, M_1; Y_1, Y_2) + n\epsilon_1 \\
&= I(M_0, M_1; Y_2) + I(M_0, M_1; Y_1|Y_2) + n\epsilon_1 \\
&\le I(M_0, M_1; Y_2) + I(M_0, M_1, Y_2; Y_1) + n\epsilon_1 \\
&\le I(M_0, M_1, W_1, Z_1; Y_2) + I(\mathcal{U}_1^1; Y_1) + n\epsilon_1 \\
&\le I(\mathcal{U}_2^3; Y_2) + I(X_1; Y_1) + n\epsilon_1.
\end{aligned} \tag{100}$$

Invoking the symmetry between receivers $Y$ and $W$, one can prove (22f) in a similar fashion.

In order to obtain the bound in (22c), we write

$$\begin{aligned}
&n(R_0 + R_1 + R_2) \le I(M_0, M_1; Y_1, Y_2) + I(M_2; Z_1, Z_2) + n\epsilon_1 + n\epsilon_2 \\
&\le I(M_0, M_1; Y_1, Y_2) + I(M_2; Z_1, Z_2|M_0, M_1) + n\epsilon_1 + n\epsilon_2 \\
&\le I(M_0, M_1; Y_2) + I(M_0, M_1; Y_1|Y_2)
\end{aligned}$$

$$+ I(M_2; Z_1, W_1|M_0, M_1) + I(M_2; Z_2|M_0, M_1, Z_1, W_1) + n\epsilon_1 + n\epsilon_2.$$

By adding and subtracting the term $I(Z_1, W_1; Y_2|M_0, M_1)$ in the above expression, we obtain

$$\begin{aligned} n(R_0 + R_1 + R_2) \leq & I(M_0, M_1, Z_1, W_1; Y_2) + I(M_0, M_1; Y_1|Y_2) + I(M_2; Z_1, W_1|M_0, M_1) \\ & + I(M_2; Z_2|M_0, M_1, Z_1, W_1) - I(Z_1, W_1; Y_2|M_0, M_1) + n\epsilon_1 + n\epsilon_2. \end{aligned} \tag{101}$$

Further bounding of the right hand side yields

$$\begin{aligned} n(R_0 + R_1 + R_2) &\leq I(\mathcal{U}_2^3; Y_2) + I(\mathcal{U}_2^2; Z_2|\mathcal{U}_2^3) + I(M_0, M_1, Y_2; Y_1) \\ &\quad + I(M_2; Z_1, W_1|M_0, M_1, Y_2) + n\epsilon_1 + n\epsilon_2 \\ &\leq I(\mathcal{U}_2^3; Y_2) + I(\mathcal{U}_2^2; Z_2|\mathcal{U}_2^3) + I(X_1; Y_1) + n\epsilon_1 + n\epsilon_2. \end{aligned}$$

In a similar manner, one can prove the bound in (22g).

In order to bound the sum rate in (22d), we have

$$\begin{aligned} & n(R_0 + R_1 + R_2 + R_3) \\ & \leq I(M_0, M_1; Y_1, Y_2) + I(M_2; Z_1, Z_2) + I(M_3; W_1, W_2) + n\epsilon_1 + n\epsilon_2 + n\epsilon_3 \\ & \leq I(M_0, M_1; Y_2) + I(M_0, M_1; Y_1|Y_2) + I(M_2; W_1, Z_1) + I(M_2; Z_2|W_1, Z_1) \\ & \quad + I(M_3; W_1, Z_1) + I(M_3; W_2|W_1, Z_1) + n\epsilon_1 + n\epsilon_2 + n\epsilon_3 \\ & \leq I(M_0, M_1; Y_2) + I(M_0, M_1; Y_1|Y_2) + I(M_2; W_1, Z_1|M_0, M_1) \\ & \quad + I(M_2; Z_2|M_0, M_1, W_1, Z_1) + I(M_3; W_1, Z_1|M_0, M_1, M_2) \\ & \quad + I(M_3; W_2|W_1, Z_1, M_0, M_1, M_2) + n\epsilon_1 + n\epsilon_2 + n\epsilon_3 \\ & = I(M_0, M_1; Y_2) + I(M_0, M_1; Y_1|Y_2) + I(M_2, M_3; Z_1, W_1|M_0, M_1) \\ & \quad + I(M_2; Z_2|M_0, M_1, W_1, Z_1) + I(M_3; W_2|W_1, Z_1, M_0, M_1, M_2) \\ & \quad + I(W_1, Z_1; Y_2|M_0, M_1) - I(W_1, Z_1; Y_2|M_0, M_1) + n\epsilon_1 + n\epsilon_2 + n\epsilon_3 \\ & \leq I(M_0, M_1, W_1, Z_1; Y_2) + I(M_0, M_1; Y_1|Y_2) \\ & \quad + I(M_2, M_3; Z_1, W_1|M_0, M_1, Y_2) + I(M_2; Z_2|M_0, M_1, W_1, Z_1) \\ & \quad + I(M_3; W_2|W_1, Z_1, M_0, M_2) + n\epsilon_1 + n\epsilon_2 + n\epsilon_3 \\ & \leq I(\mathcal{U}_2^3; Y_2) + I(\mathcal{U}_2^2; Z_2|\mathcal{U}_2^3) + I(X_2; W_2|\mathcal{U}_2^2) + I(X_1; Y_1) + n\epsilon_1 + n\epsilon_2 + n\epsilon_3. \end{aligned}$$

One can use the same technique to obtain the bound in (22h).

## Appendix X
## Proof of the inequalities in (23) and (24)

Inequality (23a) for $\text{Region}_1$ and inequality (24a) for $\text{Region}_2$ are identical, and in order to prove them, we have

$$\begin{aligned} nR_0 &\leq I(M_0; Z_1, Z_2) + n\epsilon_2 \\ &= I(M_0; Z_1) + I(M_0; Z_2|Z_1) + n\epsilon_2 \\ &\leq I(M_0, Z_2; Z_1) + I(M_0, Z_1; Z_2) + n\epsilon_2 \\ &= I(\mathcal{V}_1^3; Z_1) + I(\mathcal{V}_2^3; Z_2) + n\epsilon_2 = I(\mathcal{X}_1^3; Z_1) + I(\mathcal{X}_2^3; Z_2) + n\epsilon_2. \end{aligned}$$

We now prove (23b). Using Fano's inequality, we write

$$\begin{aligned} n(R_0 + R_2) &\leq I(M_0, M_2; Z_1, Z_2) + n\epsilon_2 \\ &= I(M_0, M_2; Z_1|Z_2) + I(M_0, M_2; Z_2) + n\epsilon_2 \\ &\leq I(M_0, M_2, Z_2; Z_1) + I(M_0, M_2, M_1, Y_1; Z_2) + n\epsilon_2 \\ &= I(\mathcal{V}_1^2; Z_1) + I(\mathcal{V}_2^2; Z_2) + n\epsilon_2. \end{aligned}$$

Similarly, to prove (24b) we have

$$\begin{aligned} n(R_0 + R_2) &\leq I(M_0, M_2; Z_1, Z_2) + n\epsilon_2 \\ &= I(M_0, M_2; Z_1) + I(M_0, M_2; Z_2|Z_1) + n\epsilon_2 \\ &\leq I(M_0, M_2, M_3, W_2; Z_1) + I(M_0, M_2, Z_1; Z_2) + n\epsilon_2 \\ &= I(\mathcal{X}_1^2; Z_1) + I(\mathcal{X}_2^2; Z_2) + n\epsilon_2. \end{aligned}$$

We now prove (23c). Using Fano's inequality we write

$$\begin{aligned} n(R_0 + R_1 + R_2) &\leq I(M_0, M_2; Z_1, Z_2) + I(M_1; Y_1, Y_2) + n\epsilon_1 + n\epsilon_2 \\ &\leq I(M_0, M_2; Z_1, Z_2) + I(M_1; Y_1, Y_2|M_0, M_2) + n\epsilon_1 + n\epsilon_2 \qquad (102) \\ &\leq I(M_0, M_2; Z_1, Z_2) + I(M_1; Y_1, Z_2|M_0, M_2) + n\epsilon_1 + n\epsilon_2 \qquad (103) \\ &= I(M_0, M_2; Z_1|Z_2) + I(M_0, M_2; Z_2) + I(M_1; Z_2|M_0, M_2) \\ &\qquad + I(M_1; Y_1|Z_2, M_0, M_2) + n\epsilon_1 + n\epsilon_2 \\ &\leq I(M_0, M_2, Z_2; Z_1) + I(M_0, M_2, M_1; Z_2) + I(M_1; Y_1|Z_2, M_0, M_2) + n\epsilon_1 + n\epsilon_2 \end{aligned}$$

$$\leq I(M_0, M_2, Z_2; Z_1) + I(M_0, M_2, M_1, Y_1; Z_2) + I(M_1; Y_1 | Z_2, M_0, M_2) + n\epsilon_1 + n\epsilon_2$$

$$= I(\mathcal{V}_1^2; Z_1) + I(\mathcal{V}_2^2; Z_2) + I(X_1; Y_1 | \mathcal{V}_1^2) + n\epsilon_1 + n\epsilon_2, \quad (104)$$

where in (102) we used the independence of $(M_0, M_2)$ and $M_1$, and in (103) we used the fact that $Z_2$ is less degraded than $Y_2$.

Using the symmetry between $(W_2, W_1)$ and $(Y_1, Y_2)$, an analogous argument can be used to prove (24c).

In order to prove (23d), we use Fano's inequality to write

$$n(R_0 + R_1 + R_2 + R_3) \leq I(M_0, M_2; Z_1, Z_2) + I(M_1; Y_1, Y_2) + I(M_3; W_1, W_2) + n\epsilon_1 + n\epsilon_2 + n\epsilon_3. \quad (105)$$

From (105) we have

$$n(R_0 + R_1 + R_2 + R_3) \leq I(M_0, M_2; Z_1, Z_2) + I(M_1; Y_1, Z_2) + I(M_3; Y_1, W_2) + n\epsilon_1 + n\epsilon_2 + n\epsilon_3 \quad (106)$$

$$\leq I(M_0, M_2; Z_1, Z_2) + I(M_1; Y_1, Z_2 | M_0, M_2) + I(M_3; Y_1, W_2 | M_0, M_2, M_1) + n\epsilon_1 + n\epsilon_2 + n\epsilon_3 \quad (107)$$

$$= I(M_0, M_2; Z_1, Z_2) + I(M_1; Y_1 | M_0, M_2) + I(M_1; Z_2 | M_0, M_2, Y_1) + I(M_3; Y_1 | M_0, M_2, M_1) + I(M_3; W_2 | M_0, M_2, M_1, Y_1) + n\epsilon_1 + n\epsilon_2 + n\epsilon_3$$

$$= I(M_0, M_2; Z_1, Z_2) + I(M_1, M_3; Y_1 | M_0, M_2) + I(M_1; Z_2 | M_0, M_2, Y_1) + I(M_3; W_2 | M_0, M_2, M_1, Y_1) + I(Z_2; Y_1 | M_0, M_2) - I(Z_2; Y_1 | M_0, M_2) + n\epsilon_1 + n\epsilon_2 + n\epsilon_3 \quad (108)$$

$$\leq I(M_0, M_2; Z_2) + I(M_0, M_2; Z_1 | Z_2) + I(M_1, M_3; Y_1 | M_0, M_2, Z_2) + I(M_3; W_2 | M_0, M_2, M_1, Y_1) + I(Z_2; Y_1 | M_0, M_2) + I(M_1; Z_2 | M_0, M_2, Y_1) + n\epsilon_1 + n\epsilon_2 + n\epsilon_3$$

$$= I(M_0, M_2, Y_1; Z_2) + I(M_0, M_2; Z_1 | Z_2) + I(M_1, M_3; Y_1 | M_0, M_2, Z_2) + I(M_3; W_2 | M_0, M_2, M_1, Y_1) + I(M_1; Z_2 | M_0, M_2, Y_1) + n\epsilon_1 + n\epsilon_2 + n\epsilon_3$$

$$\leq I(M_0, M_2, Y_1, M_1; Z_2) + I(M_0, M_2, Z_2; Z_1) + I(M_1, M_3; Y_1 | M_0, M_2, Z_2)$$

$$+ I(M_3; W_2|M_0, M_2, M_1, Y_1) + n\epsilon_1 + n\epsilon_2 + n\epsilon_3$$

$$= I(\mathcal{V}_2^1; Z_2) + I(\mathcal{V}_1^1; Z_1) + I(X_1; Y_1|\mathcal{V}_1^2) + I(X_2; W_2|\mathcal{V}_2^2) + n\epsilon_1 + n\epsilon_2 + n\epsilon_3,$$

where in (106) we have use the fact that $Y_1$ is less degraded than $W_1$, and $Z_2$ is less degraded than $Y_2$. In (107) we used the observation that $M_0$ and $M_2$ are independent of $M_1$ and $M_3$, and that $M_1$ and $M_3$ are independent of each other. The term $I(Z_2; Y_1|M_0, M_2)$ is added and subtracted in (108) in order to introduce $Z_2$ in the conditioning of the second term in (108).

Using the symmetry between $(W_2, W_1)$ and $(Y_1, Y_2)$, an analogous argument can be used to prove (24d).

## APPENDIX XI
## APPLICATION TO THE GAUSSIAN CHANNEL—REGION$_1$

In this section, we will show that Region$_1$ is an outer bound on the capacity region. In order to do this we will show that for every achievable rate vector there exist power partitions $(\alpha, \alpha', \alpha'')$ such that the inequalities in (18b)–(18i) are satisfied. The argument will be based on invoking the inequalities in (22) and (23) in the case in which each subchannel is Gaussian. (The corresponding argument for Region$_2$ is almost identical, and will be briefly discussed in Appendix XII.) We begin by observing that [2]

$$H(Y_1) \le \frac{n}{2}\log\big(2\pi e(P_1 + N_1^1)\big) \tag{109a}$$

$$H(Z_1) \le \frac{n}{2}\log\big((2\pi e(P_1 + N_1^2)\big) \tag{109b}$$

$$H(W_2) \le \frac{n}{2}\log\big((2\pi e(P_2 + N_2^1)\big) \tag{109c}$$

$$H(Z_2) \le \frac{n}{2}\log\big((2\pi e(P_2 + N_2^2)\big). \tag{109d}$$

In the following subsections we will specify the partitions $\alpha$, $\alpha'$ and $\alpha''$ and we will employ those partitions and the entropy power inequality to provide the desired bounds.

### *A. Specifying power partitions $\alpha$, $\alpha'$ and $\alpha''$*

*1) Specifying $\alpha$:* Since conditioning reduces entropy, we conclude that there exist two non-negative reals $\alpha_1^1$ and $\alpha_1^2$ satisfying $\alpha_1^1 + \alpha_1^2 \le 1$ such that

$$H(Y_1|\mathcal{U}_1^3) = \frac{n}{2}\log(2\pi e((\alpha_1^1 + \alpha_1^2)P_1 + N_1^1)), \tag{110}$$

$$H(Y_1|\mathcal{U}_1^2) = \frac{n}{2}\log(2\pi e(\alpha_1^1 P_1 + N_1^1)). \tag{111}$$

Similarly, there exist $\alpha_2^1$ and $\alpha_2^2$ satisfying $\alpha_2^1 + \alpha_2^2 \le 1$ such that

$$H(Z_2|\mathcal{U}_2^3) = \frac{n}{2}\log(2\pi e((\alpha_2^1 + \alpha_2^2)P_2 + N_2^2)), \tag{112}$$

$$H(Z_2|\mathcal{U}_2^2) = \frac{n}{2}\log(2\pi e(\alpha_2^1 P_2 + N_2^2)). \tag{113}$$

In (111) we have used the fact that $\mathcal{U}_1^2$ contains more information about $Y_1$ than $\mathcal{U}_1^3$, and that $\mathcal{U}_2^2$ contains more information about $Z_2$ than $\mathcal{U}_2^3$. This fact is immediately apparent from the definitions in (21).

*2) Specifying $\alpha'$:* Because conditioning reduces entropy, there also exist non-negative reals $\alpha_1'^1, \alpha_1'^2, \alpha_2'^1$ and $\alpha_2'^2$ such that $\alpha_i'^1 + \alpha_i'^2 \leq 1$, $i = 1, 2$, and

$$H(Y_1|M_0, Z_2) = H(Y_1|\mathcal{V}_1^3) = \log\Big(2\pi e\big((\alpha_1'^1 + \alpha_1'^2)P_1 + N_1^1\big)\Big), \tag{114}$$

$$H(W_2|M_0, Z_1) = H(W_2|\mathcal{V}_2^3) = \log\Big(2\pi e\big((\alpha_2'^1 + \alpha_2'^2)P_2 + N_2^1\big)\Big), \tag{115}$$

$$H(Y_1|M_0, M_2, Z_2) = H(Y_1|\mathcal{V}_1^2) = \log\big(2\pi e(\alpha_1'^1 P_1 + N_1^1)\big), \tag{116}$$

$$H(W_2|M_0, M_2, Y_1, M_1) = H(W_2|\mathcal{V}_2^2) = \log\big(2\pi e(\alpha_2'^1 P_2 + N_2^1)\big), \tag{117}$$

where (117) follows from the fact that $Y_1$ is a less degraded version of $Z_1$.

*3) Specifying $\alpha''$:* Using, once again, the fact that conditioning reduces entropy, one can find non-negative reals $\alpha_1''^1$, $\alpha_1''^2$, $\alpha_2''^1$, $\alpha_2''^2$ such that $\alpha_i''^1 + \alpha_i''^2 \leq 1$, $i = 1, 2$, and

$$H(Z_1|\mathcal{U}_1^3) = \frac{n}{2}\log(2\pi e((\alpha_1''^1 + \alpha_1''^2)P_1 + N_1^2)), \tag{118}$$

$$H(Z_1|\mathcal{U}_1^2) = \frac{n}{2}\log(2\pi e(\alpha_1''^1 P_1 + N_1^2)), \tag{119}$$

$$H(W_2|\mathcal{U}_2^3) = \frac{n}{2}\log(2\pi e((\alpha_2''^1 + \alpha_2''^2)P_2 + N_2^1)), \tag{120}$$

$$H(W_2|\mathcal{U}_2^2) = \frac{n}{2}\log(2\pi e(\alpha_2''^1 P_2 + N_2^1)). \tag{121}$$

### *B. Applying the entropy power inequality*

*1) Applying the entropy power inequality with $\alpha$:* Using the entropy power inequality one can show that

$$H(Z_1|\mathcal{U}_1^3) \geq \frac{n}{2}\log(2\pi e((\alpha_1^1 + \alpha_1^2)P_1 + N_1^2)), \tag{122}$$

$$H(Z_1|\mathcal{U}_1^2) \geq \frac{n}{2}\log(2\pi e(\alpha_1^1 P_1 + N_1^2)), \tag{123}$$

$$H(W_1|\mathcal{U}_1^3) \geq \frac{n}{2}\log(2\pi e((\alpha_1^1 + \alpha_1^2)P_1 + N_1^3)), \tag{124}$$

$$H(W_1|\mathcal{U}_1^2) \geq \frac{n}{2}\log(2\pi e(\alpha_1^1 P_1 + N_1^3)). \tag{125}$$

Similarly, we have

$$H(W_2|\mathcal{U}_2^3) \leq \frac{n}{2}\log(2\pi e((\alpha_2^1 + \alpha_2^2)P_2 + N_2^1)), \tag{126}$$

$$H(W_2|\mathcal{U}_2^2) \leq \frac{n}{2}\log(2\pi e(\alpha_2^1 P_2 + N_2^1)), \tag{127}$$

where in (126) and (127) we have used the entropy power inequality in the reverse direction. Using the entropy power inequality on (112) and (113), we obtain

$$H(Y_2|\mathcal{U}_2^3) \geq \frac{n}{2}\log(2\pi e((\alpha_2^1 + \alpha_2^2)P_2 + N_2^3)), \tag{128}$$

$$H(Y_2|\mathcal{U}_2^2) \geq \frac{n}{2}\log(2\pi e(\alpha_2^1 P_2 + N_2^3)). \tag{129}$$

*2) Applying the entropy power inequality with $\alpha'$:*

$$H(Z_1|M_0, Z_2) = H(Z_1|\mathcal{V}_1^3) \geq \log\Big(2\pi e\big((\alpha_1'^1 + \alpha_1'^2)P_1 + N_1^2\big)\Big), \tag{130}$$

$$H(Z_2|M_0, Z_1) = H(Z_2|\mathcal{V}_2^3) \geq \log\Big(2\pi e\big((\alpha_2'^1 + \alpha_2'^2)P_2 + N_2^2\big)\Big), \tag{131}$$

$$H(Z_1|M_0, M_2, Z_2) = H(Z_1|\mathcal{V}_1^2) \geq \log\big(2\pi e(\alpha_1'^1 P_1 + N_1^2)\big), \tag{132}$$

$$H(Z_2|M_0, M_2, Y_1, M_1) = H(Z_2|\mathcal{V}_2^2) \geq \log\big(2\pi e(\alpha_2'^1 P_2 + N_2^2)\big). \tag{133}$$

*3) Applying the entropy power inequality with $\alpha''$:*

$$H(Y_1|\mathcal{U}_1^3) \leq \frac{n}{2}\log(2\pi e((\alpha_1''^1 + \alpha_1''^2)P_1 + N_1^1)), \tag{134}$$

$$H(W_1|\mathcal{U}_1^3) \geq \frac{n}{2}\log(2\pi e((\alpha_1''^1 + \alpha_1''^2)P_1 + N_1^3)), \tag{135}$$

$$H(Y_1|\mathcal{U}_1^2) \leq \frac{n}{2}\log(2\pi e(\alpha_1''^1 P_1 + N_1^1)), \tag{136}$$

$$H(W_1|\mathcal{U}_1^2) \geq \frac{n}{2}\log(2\pi e(\alpha_1''^1 P_1 + N_1^3)), \tag{137}$$

$$H(Z_2|\mathcal{U}_2^3) \geq \frac{n}{2}\log(2\pi e((\alpha_2''^1 + \alpha_2''^2)P_2 + N_2^2)), \tag{138}$$

$$H(Y_2|\mathcal{U}_2^3) \geq \frac{n}{2}\log(2\pi e((\alpha_2''^1 + \alpha_2''^2)P_2 + N_2^3)), \tag{139}$$

$$H(Z_2|\mathcal{U}_2^2) \geq \frac{n}{2}\log(2\pi e(\alpha_2''^1 P_2 + N_2^2)), \tag{140}$$

$$H(Y_2|\mathcal{U}_2^2) \geq \frac{n}{2}\log(2\pi e(\alpha_2''^1 P_2 + N_2^3)). \tag{141}$$

Using (109a)–(141) we now prove our target inequalities.

### *C. Proving the converse of the inequalities in* (18b)–(18c)

For this set of inequalities, we will apply the inequalities in (109), (110)–(113) and (122)–(129) to (22a)–(22d). As in (14), we will use $C(x)$ to denote $\frac{1}{2}\log(1+x)$.

*1) Proving the converse of the first inequality in* (18b)*:* From (22a), we have

$$\begin{aligned}nR_0 &\leq H(Y_1) - H(Y_1|\mathcal{U}_1^3) + H(Y_2) - H(Y_2|\mathcal{U}_1^3) + n\epsilon_1\\&\leq \frac{n}{2}\log(2\pi e(P_1+N_1^1)) - \frac{n}{2}\log(2\pi e((\alpha_1^1+\alpha_1^2)P_1+N_1^1)) + \frac{n}{2}\log(2\pi e(P_2+N_2^3))\\&\quad - \frac{n}{2}\log(2\pi e((\alpha_2^1+\alpha_2^2)P_2+N_2^3)) + n\epsilon_1\\&= nC\Big(\frac{\alpha_1^3 P_1}{N_1^1+(\alpha_1^1+\alpha_1^2)P_1}\Big) + nC\Big(\frac{\alpha_2^3 P_2}{N_2^3+(\alpha_2^2+\alpha_2^1)P_2}\Big) + n\epsilon_1.\end{aligned}$$

*2) Proving the converse of the second inequality in* (18b)*:* From (22b), we have

$$\begin{aligned}n(R_0+R_1) &\leq H(Y_1) - H(Y_1|X_1) + H(Y_2) - H(Y_2|\mathcal{U}_2^3) + n\epsilon_1\\&\leq \frac{n}{2}\log(2\pi e(P_1+N_1^1)) - \frac{n}{2}\log(2\pi e(N_1^1)) + \frac{n}{2}\log(2\pi e(P_2+N_2^3))\\&\quad - \frac{n}{2}\log(2\pi e((\alpha_2^1+\alpha_2^2)P_2+N_2^3)) + n\epsilon_1\\&= nC\Big(\frac{P_1}{N_1^1}\Big) + nC\Big(\frac{\alpha_2^3 P_2}{N_2^3+(\alpha_2^2+\alpha_2^1)P_2}\Big) + n\epsilon_1.\end{aligned}$$

*3) Proving the converse of the third inequality in* (18b)*:* Using (22c), we have

$$\begin{aligned}n(R_0+R_1+R_2) &\leq I(\mathcal{U}_2^3;Y_2) + I(\mathcal{U}_2^2;Z_2|\mathcal{U}_2^3) + I(X_1;Y_1) + n\epsilon_1 + n\epsilon_2\\&\leq H(Y_1) - H(Y_1|X_1) + H(Y_2) - H(Y_2|\mathcal{U}_2^3) + H(Z_2|\mathcal{U}_2^3)\\&\qquad - H(Z_2|\mathcal{U}_2^2) + n\epsilon_1 + n\epsilon_2\\&\leq nC\Big(\frac{P_1}{N_1^1}\Big) + nC\Big(\frac{\alpha_2^3 P_2}{N_2^3+(\alpha_2^2+\alpha_2^1)P_2}\Big)\\&\qquad + \frac{n}{2}\log(2\pi e((\alpha_2^1+\alpha_2^2)P_2+N_2^2)) - \frac{n}{2}\log(2\pi e(\alpha_2^1 P_2+N_2^2))\\&\qquad + n\epsilon_1 + n\epsilon_2\\&= nC\Big(\frac{P_1}{N_1^1}\Big) + nC\Big(\frac{\alpha_2^3 P_2}{N_2^3+(\alpha_2^2+\alpha_2^1)P_2}\Big) + nC\Big(\frac{\alpha_2^2 P_2}{N_2^2+\alpha_2^1 P_2}\Big)\\&\qquad + n\epsilon_1 + n\epsilon_2.\end{aligned}$$

*4) Proving the converse of the inequality in* (18c)*:* To prove the converse of this inequality we use (22d) to write

$$\begin{aligned}n(R_0+R_1+R_2+R_3) &\leq nC\Big(\frac{P_1}{N_1^1}\Big) + nC\Big(\frac{\alpha_2^3 P_2}{N_2^3+(\alpha_2^2+\alpha_2^1)P_2}\Big) + nC\Big(\frac{\alpha_2^2 P_2}{N_2^2+\alpha_2^1 P_2}\Big)\\&\qquad + H(W_2|\mathcal{U}_2^2) - H(W_2|X_2) + n\epsilon_1 + n\epsilon_2 + n\epsilon_3\\&\leq nC\Big(\frac{P_1}{N_1^1}\Big) + nC\Big(\frac{\alpha_2^3 P_2}{N_2^3+(\alpha_2^2+\alpha_2^1)P_2}\Big) + nC\Big(\frac{\alpha_2^2 P_2}{N_2^2+\alpha_2^1 P_2}\Big)\end{aligned}$$

$$+\frac{n}{2}\log(2\pi e(\alpha_2^1P_2+N_2^1))-\frac{n}{2}\log(2\pi eN_2^1)$$
$$+n\epsilon_1+n\epsilon_2+n\epsilon_3, \tag{142}$$

where in (142) we have used the upper bound on $H(W_2|\mathcal{U}_2^2)$ in (127) and the fact that $H(W_2|X_2)=\frac{n}{2}\log(2\pi eN_2^1)$.

*D. Proving the converse of the inequalities in* (18d)–(18e)

For this set of inequalities, we will apply the inequalities in (109), (114)–(115) and (130)–(133) to (23a)–(23d).

*1) Proving the converse of the first inequality in* (18d)*:* Expressing (23a) in terms of the conditional entropy and using (130) and (131), one can show that

$$nR_0\le nC\Big(\frac{\alpha_1'^3P_1}{N_1^2+(\alpha_1^1+\alpha_1^2)P_1}\Big)+nC\Big(\frac{\alpha_2'^3P_2}{N_2^2+(\alpha_2^1+\alpha_2^2)P_2}\Big)+n\epsilon_2, \tag{143}$$

where $\alpha_i'^3=1-(\alpha_i'^1+\alpha_i'^2), i=1,2$.

*2) Proving the converse of the second inequality in* (18d)*:* Expressing (23b) in terms of the conditional entropy and using (109), (132) and (133), we have

$$n(R_0+R_2)\le nC\Big(\frac{(\alpha_2'^2+\alpha_2'^3)P_2}{N_2^2+\alpha_2'^1P_2}\Big)+nC\Big(\frac{(\alpha_1'^2+\alpha_1'^3)P_1}{N_1^2+\alpha_1'^1P_1}\Big)+n\epsilon_2. \tag{144}$$

*3) Proving the converse of the third inequality in* (18d)*:* Using (23c), we have

$$n(R_0+R_1+R_2)\le H(Z_2)-H(Z_2|M_0,M_2,Y_1,M_1)+H(Z_1)-H(Z_1|M_0,M_2,Z_2)$$
$$+H(Y_1|M_0,M_2,Z_2)-H(Y_1|M_1,M_3,M_0,M_2,Z_2)+n(\epsilon_1+\epsilon_2) \tag{145}$$
$$\le\frac{n}{2}\log\big((2\pi e(P_2+N_2^2)\big)-\frac{n}{2}\log\big(2\pi e(\alpha_2'^1P_2+N_2^2\big)+\frac{n}{2}\log\big((2\pi e(P_1+N_1^2)\big)$$
$$-\frac{n}{2}\log\big(2\pi e(\alpha_1'^1P_1+N_1^2)\big)+\frac{n}{2}\log\big(2\pi e(\alpha_1'^1P_1+N_1^1)\big)-\frac{n}{2}\log(2\pi eN_1^1)+n(\epsilon_1+\epsilon_2)$$
$$=nC\Big(\frac{(\alpha_2'^2+\alpha_2'^3)P_2}{N_2^2+\alpha_2'^1P_2}\Big)+nC\Big(\frac{(\alpha_1'^2+\alpha_1'^3)P_1}{N_1^2+\alpha_1'^1P_1}\Big)+nC\Big(\frac{\alpha_1'^1P_1}{N_1^1}\Big)+n(\epsilon_1+\epsilon_2). \tag{146}$$

*4) Proving the converse of the inequality in* (18e)*:* Using (23d), we have

$$n(R_0+R_1+R_2+R_3)\le H(Z_2)-H(Z_2|M_0,M_2,M_1,Y_1)+H(Z_1)-H(Z_1|M_0,M_2,Z_2)+H(Y_1|M_0,M_2,Z_2)$$
$$-H(Y_1|M_0,M_2,Z_2,M_1,M_3)+H(W_2|M_0,M_2,Y_1,M_1)$$
$$-H(W_2|M_0,M_2,Y_1,M_1,M_3)+n(\epsilon_1+\epsilon_2+\epsilon_3)$$
$$\le\frac{n}{2}\log\big((2\pi e(P_2+N_2^2)\big)-\frac{n}{2}\log\big(2\pi e(\alpha_2'^1P_2+N_2^2\big)+\frac{n}{2}\log\big(2\pi e(P_1+N_1^2)\big)$$

$$
\begin{aligned}
&\qquad - \frac{n}{2}\log\big(2\pi e(\alpha_1'^1 P_1 + N_1^2)\big) \\
&+ \frac{n}{2}\log\big(2\pi e(\alpha_1'^1 P_1 + N_1^1)\big) - H(Y_1|M_0, M_2, Z_2, M_1, M_3) + \frac{n}{2}\log\big(2\pi e(\alpha_2'^1 P_2 + N_2^1)\big) \\
&\qquad - H(W_2|M_0, M_2, Y_1, M_1, M_3) + n(\epsilon_1 + \epsilon_2 + \epsilon_3) \\
= {}& nC\Big(\frac{(\alpha_2'^2 + \alpha_2'^3)P_2}{N_2^2 + \alpha_2'^1 P_2}\Big) + nC\Big(\frac{(\alpha_1'^2 + \alpha_1'^3)P_1}{N_1^2 + \alpha_1'^1 P_1}\Big) \\
&+ nC\Big(\frac{\alpha_1'^1 P_1}{N_1^1}\Big) + nC\Big(\frac{\alpha_2'^1 P_2}{N_2^1}\Big) + n(\epsilon_1 + \epsilon_2 + \epsilon_3), \qquad (147)
\end{aligned}
$$

where in (147) we have used the fact that

$$H(Y_1|M_0, M_2, Z_2, M_1, M_3) \geq \log(2\pi e N_1^1),$$

and

$$H(W_2|M_0, M_2, Y_1, M_1, M_3) \geq \log(2\pi e N_2^1).$$

### *E. Proving the converse of the inequalities in* (18f)–(18g)

For this set of inequalities we will apply the inequalities in (109), (118)–(121), (134)–(141) to (22e)–(22h).

*1) Proving the converse of the first inequality in* (18f)*:*

$$
\begin{aligned}
nR_0 &\leq H(W_1) - H(W_1|\mathcal{U}_1^3) + H(W_2) - H(W_2|\mathcal{U}_2^3) + n\epsilon_3 \\
&\leq \frac{n}{2}\log(2\pi e(P_1 + N_1^3)) - \frac{n}{2}\log(2\pi e(N_1^3 + (\alpha_1''^1 + \alpha_1''^2)P_1)) + \frac{n}{2}\log(2\pi e(P_2 + N_2^1)) \\
&\qquad - \frac{n}{2}\log(2\pi e((\alpha_2''^1 + \alpha_2''^2)P_2 + N_2^1)) + n\epsilon_3 \\
&= nC\Big(\frac{\alpha_1''^3 P_1}{N_1^3 + (\alpha_1''^2 + \alpha_1''^1)P_1}\Big) + nC\Big(\frac{\alpha_2''^3 P_2}{N_2^1 + (\alpha_2''^2 + \alpha_2''^1)P_2}\Big) + n\epsilon_3.
\end{aligned}
$$

*2) Proving the converse of the second inequality in* (18f)*:* Using (22f) we have

$$
\begin{aligned}
n(R_0 + R_3) &\leq H(W_2) - H(W_2|X_2) + H(W_1) - H(W_1|\mathcal{U}_1^3) + n\epsilon_3 \\
&\leq \frac{n}{2}\log(2\pi e(P_2 + N_2^1)) - \frac{n}{2}\log(2\pi e N_2^1) + \frac{n}{2}\log(2\pi e(P_1 + N_1^3)) \\
&\qquad - \frac{n}{2}\log(2\pi e((\alpha_1''^1 + \alpha_1''^2)P_1 + N_1^3)) + n\epsilon_3 \\
&= nC\Big(\frac{P_2}{N_2^1}\Big) + nC\Big(\frac{\alpha_1''^3 P_1}{N_1^3 + (\alpha_1''^1 + \alpha_1''^2)P_1}\Big) + n\epsilon_3.
\end{aligned}
$$

*3) Proving the converse of the third inequality in* (18f)*:* In order to prove this inequality we have from (22g)

$$n(R_0+R_2+R_3)) \leq H(W_1)-H(W_1|\mathcal{U}_1^3)+H(Z_1|\mathcal{U}_1^3)-H(Z_1|\mathcal{U}_1^2)$$
$$+H(W_2)-H(W_2|X_2)+n\epsilon_2+n\epsilon_3 \tag{148}$$
$$\leq \frac{n}{2}\log(2\pi e(P_1+N_1^3))-\frac{n}{2}\log(2\pi e((\alpha_1''^1+\alpha_1''^2)P_1+N_1^3))$$
$$+\frac{n}{2}\log(2\pi e((\alpha_1''^1+\alpha_1''^2)P_1+N_1^2))-\frac{n}{2}\log(2\pi e(\alpha_1''^1P_1+N_1^2))$$
$$+\frac{n}{2}\log(2\pi e(P_2+N_2^1))-\frac{n}{2}\log(2\pi eN_2^1)+n\epsilon_2+n\epsilon_3 \tag{149}$$
$$\leq nC\Big(\frac{\alpha_1''^3P_1}{N_1^3+(\alpha_1''^2+\alpha_1''^1)P_1}\Big)+nC\Big(\frac{\alpha_1''^2P_1}{N_1^2+\alpha_1''^1P_1}\Big)$$
$$+nC\Big(\frac{P_2}{N_2^1}\Big)+n\epsilon_2+n\epsilon_3. \tag{150}$$

*4) Proving the converse of the inequality in* (18g)*:* Using (22h), we have

$$n(R_0+R_1+R_2+R_3) \leq nC\Big(\frac{\alpha_1''^3P_1}{N_1^3+(\alpha_1''^2+\alpha_1''^1)P_1}\Big)+nC\Big(\frac{\alpha_1''^2P_1}{N_1^2+\alpha_1''^1P_1}\Big)$$
$$+nC\Big(\frac{P_2}{N_2^1}\Big)+H(Y_1|\mathcal{U}_1^2)-H(Y_1|X_1)+n\epsilon_1+n\epsilon_2+n\epsilon_3$$
$$\leq nC\Big(\frac{\alpha_1''^3P_1}{N_1^3+(\alpha_1''^2+\alpha_1''^1)P_1}\Big)+nC\Big(\frac{\alpha_1''^2P_1}{N_1^2+\alpha_1''^1P_1}\Big)$$
$$+nC\Big(\frac{P_2}{N_2^1}\Big)+\frac{n}{2}\log(2\pi e(\alpha_1''^1P_1+N_1^1))-\frac{n}{2}\log(2\pi eN_1^1)$$
$$+n\epsilon_1+n\epsilon_2+n\epsilon_3$$
$$= nC\Big(\frac{\alpha_1''^3P_1}{N_1^3+(\alpha_1''^2+\alpha_1''^1)P_1}\Big)+nC\Big(\frac{\alpha_1''^2P_1}{N_1^2+\alpha_1''^1P_1}\Big)$$
$$+nC\Big(\frac{P_2}{N_2^1}\Big)+nC\Big(\frac{\alpha_1''^1P_1}{N_1^1}\Big)+n\epsilon_1+n\epsilon_2+n\epsilon_3.$$

## APPENDIX XII
## APPLICATION TO THE GAUSSIAN CHANNEL—REGION$_2$

The application of the entropy power inequality to show the maximality of Region$_2$ uses essentially the same methodology as that used in Appendix XI for Region$_1$, but with the partitions $\alpha'$ chosen so as to satisfy the following equalities

$$H(Y_1|\mathcal{X}_1^3)=\frac{n}{2}\log\Big(2\pi e\big((\alpha_1'^1+\alpha_1'^2)P_1+N_1^1\big)\Big), \tag{151}$$
$$H(W_2|\mathcal{X}_2^3)=\frac{n}{2}\log\Big(2\pi e\big((\alpha_2'^1+\alpha_2'^2)P_2+N_2^1\big)\Big), \tag{152}$$

$$H(Y_1|\mathcal{X}_1^2) = \frac{n}{2}\log\big(2\pi e(\alpha_1'^1 P_1 + N_1^1)\big), \tag{153}$$

$$H(W_2|\mathcal{X}_2^2) = \frac{n}{2}\log\big(2\pi e(\alpha_2'^1 P_2 + N_2^1)\big). \tag{154}$$

Using these partitions along with the inequalities in (24a)–(24d) yields

$$n(R_0 + R_2) \le nC\Big(\frac{(\alpha_2'^2 + \alpha_2'^3)P_2}{N_2^2 + \alpha_2'^1 P_2}\Big) + nC\Big(\frac{(\alpha_1'^2 + \alpha_1'^3)P_1}{N_1^2 + \alpha_1'^1 P_1}\Big) + n\epsilon_2 \tag{155}$$

$$n(R_0 + R_2 + R_3) \le nC\Big(\frac{(\alpha_2'^2 + \alpha_2'^3)P_2}{N_2^2 + \alpha_2'^1 P_2}\Big) + nC\Big(\frac{(\alpha_1'^2 + \alpha_2'^3)P_1}{N_1^2 + \alpha_1'^1 P_1}\Big) + nC\Big(\frac{\alpha_2'^1 P_2}{N_2^1}\Big) + n(\epsilon_2 + \epsilon_3) \tag{156}$$

$$n(R_0 + R_1 + R_2 + R_3) \le nC\Big(\frac{(\alpha_2'^2 + \alpha_2'^3)P_2}{N_2^2 + \alpha_2'^1 P_2}\Big) + nC\Big(\frac{(\alpha_1'^2 + \alpha_2'^3)P_1}{N_1^2 + \alpha_1'^1 P_1}\Big) + nC\Big(\frac{\alpha_1'^1 P_1}{N_1^1}\Big) + C\Big(\frac{\alpha_2'^1 P_2}{N_2^1}\Big) + n(\epsilon_1 + \epsilon_2 + \epsilon_3), \tag{157}$$

which is the desired converse.

## Acknowledgement

The authors would like to express their gratitude to Dr. Zhi-Quan (Tom) Luo of the University of Minnesota for his insightful comments and suggestions during the development of this work.